\documentclass[a4paper,10pt]{article}

\usepackage[utf8]{inputenc}
\usepackage[T1]{fontenc}

\usepackage{a4wide}

\usepackage{amsmath,amssymb}
\usepackage{amsthm}
\usepackage{enumerate} 

\usepackage{authblk}

\usepackage[dvipsnames,table,xcdraw,svgnames]{xcolor}
\definecolor{highlightNEW}{named}{black}

\usepackage{hyperref}
\hypersetup{
    breaklinks=true,
    colorlinks=true,
    linkcolor=Navy,
    citecolor=Navy,
    urlcolor=Navy
}

\usepackage{graphicx}
\graphicspath{{fig/}}
\usepackage{epstopdf}
\usepackage[section]{placeins} 
\usepackage{subfig}

\usepackage{booktabs}
\usepackage{multirow}

\usepackage[absolute,overlay,showboxes]{textpos}
\TPMargin{2mm}
\textblockrulecolour{Navy}

\usepackage[round]{natbib}
\bibpunct[, ]{(}{)}{;}{a}{}{,}

\usepackage{soulutf8}
\sethlcolor{LimeGreen}

\soulregister\cite7%
\soulregister\citep7%
\soulregister\eqref7%
\soulregister\pageref7%
\soulregister\ref7%

\newtheorem{theorem}{Theorem}[section] 
\newtheorem{algorithm}[theorem]{Algorithm}

\newcommand{\doi}[1]{DOI~\href{\detokenize{http://dx.doi.org/#1}}{\detokenize{#1}}}
\newcommand{\zblnumber}[1]{Zbl~\href{\detokenize{https://zbmath.org/?q=an:#1}}{\detokenize{#1}}}
\newcommand{\mrnumber}[1]{\href{\detokenize{https://www.ams.org/mathscinet-getitem?mr=#1}}{\detokenize{MR#1}}}

\newcommand{\BS}{\operatorname{BS}}
\newcommand{\Cov}{\operatorname{Cov}}
\renewcommand{\d}{\,\mathrm{d}}
\newcommand{\e}{\mathrm{e}}
\newcommand{\E}{\mathbb{E}}
\newcommand{\F}{\mathcal{F}}

\newcommand{\N}{\mathbb{N}}
\newcommand{\Nn}{\mathcal{N}}
\newcommand{\Oo}{\mathcal{O}}

\newcommand{\Q}{Q}

\newcommand{\Var}{\operatorname{Var}}

\newcommand\abs[1]{\left|#1\right|} 

\newdimen\CdotAxis
\newcommand*{\CdotAux}[3]{%
  {%
    \settoheight\CdotAxis{$#2\vcenter{}$}%
    \sbox0{%
      \raisebox\CdotAxis{%
        \scalebox{#1}{%
          \raisebox{-\CdotAxis}{%
            $\mathsurround=0pt #2#3$%
          }%
        }%
      }%
    }%
    \dp0=0pt %
    \sbox2{$#2\bullet$}%
    \ifdim\ht2<\ht0 %
      \ht0=\ht2 %
    \fi
    \sbox2{$\mathsurround=0pt #2#3$}%
    \hbox to \wd2{\hss\usebox{0}\hss}%
  }%
}
\makeatletter
\def\mathcolor#1#{\@mathcolor{#1}}
\def\@mathcolor#1#2#3{%
  \protect\leavevmode
  \begingroup
    \color#1{#2}#3%
  \endgroup
}
\makeatother

\newcommand{\ccode}[2]{\par
        \vspace*{8pt}
        {{\leftskip18pt\rightskip\leftskip
        \noindent{\it #1}\/: #2\par}}\par}
\newcommand{\keywords}[1]{\ccode{Keywords}{#1}}
\newcommand{\email}[1]{\href{mailto:#1}{#1}}

\def\received#1{Received~#1\par}
\def\revised#1{Revised~#1\par}

\usepackage[mathscr]{euscript}
\DeclareSymbolFont{rsfs}{U}{rsfs}{m}{n}
\DeclareSymbolFontAlphabet{\mathscrsfs}{rsfs}

\newcommand{\jpTitle}{On simulation of rough Volterra stochastic volatility models}
\newcommand{\jpAuthors}{J. Matas and J. Posp\'{\i}\v{s}il}
\newcommand{\jpKeywords}{Volterra stochastic volatility; rough volatility; rough Bergomi model; fractional Brownian motion}
\newcommand{\jpMSC}{60G22; 65C05; 91G60}
\newcommand{\jpJEL}{C63; G12}
\newcommand{\jpDateReceived}{26 July 2021} 
\newcommand{\jpDateRevised}{2 August 2022}
\newcommand{\jpDate}{}

\author[1]{Jan Matas} 
\author[1]{Jan Posp\'{\i}\v{s}il\thanks{Corresponding author, \email{honik@kma.zcu.cz}}} 
\affil[1]{NTIS - New Technologies for the Information Society, Faculty of Applied Sciences, \authorcr University of West Bohemia, Univerzitn\'{\i} 2732/8, 301 00 Plze\v{n}, Czech Republic,\vspace*{-\baselineskip}}

\ifpdf
\hypersetup{
  pdftitle={\jpTitle},
  pdfauthor={\jpAuthors},
  pdfkeywords={\jpKeywords},
  pdfinfo={
      MSC={\jpMSC},
      JEL={\jpJEL}
  }
}
\fi

\title{\textcolor{Navy}{\textsc{\jpTitle}}}
\date{\jpDate}

\begin{document}

\maketitle

\begin{center}
\received{\jpDateReceived}
\revised{\jpDateRevised}
\end{center}

\begin{abstract}
Rough Volterra volatility models are a progressive and promising field of research in derivative pricing. Although rough fractional stochastic volatility models already proved to be superior in real market data fitting, techniques used in simulation of these models are still inefficient in terms of speed and accuracy. This paper aims to present accurate and efficient tools and techniques for Monte-Carlo simulations for a wide range of rough volatility models. In particular, we compare three commonly used simulation methods: the Cholesky method, the Hybrid scheme, and the rDonsker scheme. We also comment on the implementation of variance reduction techniques. In particular, we show the obstacles of the so-called turbocharging technique whose performance is sometimes counter-productive. To overcome these obstacles, we suggest several modifications.

\end{abstract}

\ccode{MSC classification}{\jpMSC}
\ccode{JEL classification}{\jpJEL}

\setcounter{tocdepth}{2}
\tableofcontents

\section{Introduction}\label{sec:introduction}

In mathematical finance, it is well known that one of the main issues of the Black-Scholes model lies in its assumptions about volatility of the modeled instrument's underlying asset. Opposed to the model assumptions, the realized volatility time series tends to cluster depending on the spot asset level and it certainly does not take on a constant value within a reasonable time-frame \cite{Cont01}. To deal with such aforementioned inconsistencies, stochastic volatility (SV) models were proposed originally by \citet{HullWhite87} and later e.g. by \citet{Heston93}. These models do not only assume that the asset price follows a specific stochastic process, but also that the instantaneous volatility of asset returns is of random nature as well. Specifically, the latter approach by Heston became popular in the eyes of both practitioners and academics. Several modifications of this model have been proposed over the last 20 years, see for example the literature review in \cite{MerinoPospisilSobotkaSottinenVives21ijtaf}.

Although many SV models have been proposed since the original \citet{HullWhite87} model, it seems that none of them can be considered as the universal best market practice approach. Several models might perform well when calibrated to describe complex volatility surfaces, but can suffer from over-fitting or they might not be robust in the sense described in \cite{PospisilSobotkaZiegler19ee}. Also, a model with a good fit to an implied volatility surface might not be in-line with the observed time-series properties. Independent increments of the Brownian motion turned out to be such a one severe limitation. This helped to boost the popularity of the fractional Brownian motion (fBm), a generalization of the Brownian motion, which allows correlation of increments depending on the so-called Hurst index $H \in (0,1)$.

The pioneers of the fractional SV models -- \citet{ComteRenault98}, \cite{Comte12} -- considered the Hurst parameter $H \in (1/2,1)$ which implies that the spot variance evolution is represented by a persistent process, i.e. it would have the long-memory property. In \cite{Alos07}, a mean-reverting fractional stochastic volatility model with $H\in (0,1)$ was presented. \cite{Gatheral18,Bayer16} came up with a more detailed analysis of rough fractional volatility models that should be consistent with market option prices, with realized volatility time series, and also provide superior volatility prediction results to several other models \citep{Bennedsen17}. Several approaches to the exact and approximate option pricing in models where volatility is a fractional Ornstein--Uhlenbeck (fOU) or fractional Cox--Ingersoll--Ross process were introduced by \cite{Mishura2019fOU}. Only very recently, \cite{Wang2021} applied fOU to model the logarithmic daily realized volatilities. An approach considering a two-factor fractional volatility model, combining a rough term ($H<{1}/{2}$) and a persistent term ($H>{1}/{2}$), was presented in \cite{Funahashi17}. An idea of decoupling the short- and long-term behavior of SV has ben recently developed by \cite{Bennedsen21}. 

Since the well-known monographs on numerical methods for stochastic differential equations (SDEs) by \cite{KloedenPlaten92,Milstein95,MilsteinTretyakov04}, numerical SDEs with colored or fractional noise have gained popularity, and there have been several papers available. For numerical methods for SDEs with color noise, books by \cite{LeMaitre10} or \cite{Xiu10} might be considered as a good starting point. In order to simulate SDEs with fractional noise, one can consider for example a natural fractional modification of the classical Euler-Maruyama and Milstein methods \citep{Deya12} including multi-level Monte-Carlo methods \citep{Kloeden11}. Similar approach has been applied to stochastic Volterra equations only very recently, see e.g. preprints by \cite{Li20} and \cite{Richard20}. However, it is worth to mention that not all techniques for simulating fBm \citep{Dieker02} are suitable for simulation of the fractional SV models. 

In this paper, we consider the $\alpha$RFSV model recently introduced by \cite{MerinoPospisilSobotkaSottinenVives21ijtaf}. This model unifies and generalizes the non-stationary RFSV model ($\alpha = 1$) and the rBergomi model ($\alpha = 0$). For the pricing of European-type options, we employ Monte-Carlo (MC) simulations. We compare three simulation methods: the Cholesky method (exact method), the Hybrid scheme, and the rDonsker scheme (both are approximate methods). We show that all the three methods are appropriate for the simulation of the model in terms of accuracy. We then compare the speed of the three methods.  We also implement a variance reduction method referred to as turbocharging \citep{McCrickerd18} and analyze its effect on the variance in price estimations. We believe that its importance is somewhat overestimated in the literature and we show on specific examples that it does not always work well. As a solution, we propose a simple modification to overcome theses obstacles.

The structure of the paper is the following. In Section~\ref{sec:preliminaries}, we introduce the studied rough Volterra stochastic volatility models. 
In Section~\ref{sec:methods}, we describe the methodology, in particular the details about Monte-Carlo simulations techniques used (Sections \ref{s:meth:Simul} and \ref{s:meth:Simul:SimMethods}), as well as the suitability of the so called turbocharging variance reduction technique (Section~\ref{s:meth:Simul:turbo}).
In Section~\ref{s:result:simul}, we present numerical results and compare all the three considered methods. Then, we provide results for our modified variance reduction techniques.
We conclude all obtained results in Section~\ref{sec:conclusion}.

\section{Preliminaries and notation}\label{sec:preliminaries}

\subsection{Rough Volterra volatility models}

Let $S=(S_{t}, t\in[0,T])$ be a strictly positive asset price process under a market-chosen risk-neutral probability measure $\Q $ that can be represented as
\begin{eqnarray}\label{e:model}
\d S_{t}= r S_{t} \d t + \sigma_{t}S_{t}  \left(\rho \d W_{t}  + \sqrt{1-\rho^{2}}\d \widetilde{W}_{t} \right),
\end{eqnarray}
where $S_{0}$ is the current spot price, $r\geq 0$ is the interest rate, $W_{t}$ and $\widetilde{W}_{t}$ are independent standard Wiener processes defined on a probability space $(\Omega, \mathcal{F}, \Q)$ and $\rho\in[-1,1]$. Also, recall that for any $\rho\in[-1,1]$, a process $\rho W_t + \sqrt{1-\rho^{2}}\, \widetilde{W}_t$ is also a standard Wiener process. Let $\mathcal{F}_t^{W}$ and $\mathcal{F}_t^{\widetilde{W}}$ be the two filtrations generated by $W_t$ and $\widetilde{W}_t$ respectively and let $\mathcal{F}_t:=\mathcal{F}_t^{W} \cup \mathcal{F}_t^{\widetilde{W}}$ (for each $t\geq 0$, $\mathcal{F}_t$ is the minimal sigma algebra that includes both sigma algebras $\mathcal{F}_t^{W}$ and $\mathcal{F}_t^{\widetilde{W}}$).

The \emph{stochastic volatility process} $\sigma_{t}$ is a square-integrable process, adapted to the filtration generated by $W_t$ whose trajectories are assumed to be a.s. c\`adl\`ag and stricly positive a.e. For convenience, we let $X_{t} = \ln S_{t}$, $t\in[0,T]$ which lead to the differential representation
\begin{eqnarray}\label{e:log-model}
\d X_{t}= \left(r -\frac{1}{2}\sigma^{2}_{t}\right)\d t + \sigma_{t} \left(\rho \d W_{t}  + \sqrt{1-\rho^{2}} \d \widetilde{W}_{t} \right).
\end{eqnarray}

From now on, we consider the model represented by Equation \eqref{e:log-model} with \emph{general Volterra volatility process} defined as
\begin{equation} \label{e:volprocess}
\sigma_t := f(t, Y_t), \quad t\geq 0,
\end{equation}
where $f: [0,+\infty) \times \mathbb{R} \mapsto [0,+\infty)$ is a deterministic function such that $\sigma_t$ belongs to $L^{1}(\Omega \times [0,+\infty))$ and $Y = (Y_t, t\geq 0)$ is the Gaussian Volterra process
\begin{equation} \label{e:Volterra}
Y_t = \int_0^t K(t,s)\,\d W_s,
\end{equation}
where $K(t,s)$ is a kernel such that for all $t>0$
\begin{equation}
\int\limits_0^t K^2(t,s) \d s < \infty \tag{A1}\label{A1}
\end{equation}
and
\begin{equation}
\F^Y_t = \F^W_t. \tag{A2}\label{A2}
\end{equation}

Next, denote the autocovariance function of $Y_t$ by $r(t,s)$ and the variance of $Y_t$ by $r(t)$, i.e.:
\begin{align}
r(t,s) &:= \E[Y_t Y_s], \quad t,s\geq 0, \notag \\
r(t) &:= r(t,t) = \E[Y_t^2], \quad t\geq 0. \label{e:r}
\end{align}

In particular we assume that $X_t$ is the log-price process \eqref{e:log-model} with $\sigma_t$ being the \emph{exponential Volterra volatility process}
\begin{equation}\label{e:expVolterra}
\sigma_t = f(t,Y_t) = \sigma_{0}\exp\left\{\xi Y_t {- \frac12 \alpha\xi^2 r(t)} \right\}, \quad t\geq 0,
\end{equation}
where $(Y_t,t\geq 0)$ is the Gaussian Volterra process \eqref{e:Volterra} satisfying assumptions \eqref{A1} and \eqref{A2}, $r(t)$ is its autocovariance function \eqref{e:r}, and $\sigma_0>0$, $\xi>0$, and $\alpha\in[0,1]$ are model parameters.

Let us now focus on a very important example of a Gaussian Volterra process, namely the \emph{standard fractional Brownian motion} (fBm) $B^{H}_{t}$, which can be represented by
\begin{equation} \label{e:fBmV}
B_t^H = \int_0^t K(t,s)\,\d W_s,
\end{equation}
where $K(t,s)$ is a kernel that depends also on the Hurst parameter $H \in (0,1)$. Recall that the autocovariance function of $B^{H}_{t}$ is given by
\begin{equation}\label{e:fBmCov}
r(t,s):=\E[B^H_t B^H_s] = \frac12 \left( t^{2H} + s^{2H} - |t-s|^{2H}\right), \quad t,s\geq 0,
\end{equation}
and in particular $r(t):=r(t,t) = t^{2H}$, $t\geq0$.

It is well known that fBm has several different integral representations. The socalled Riemann-Liouville representation was introduced already by \cite{Levy53} and it takes the form
\begin{equation}\label{e:fBmRL}
B_t^H  = \frac1{\Gamma(H+1/2)} \int_0^t (t-s)^{H-1/2}\,\d W_s.
\end{equation}
Its \emph{ill-suitability for applications} \citep[p. 424]{Mandelbrot68} lead to a \emph{more suitable}, but semi-infinite domain representation
\begin{equation}\label{e:fBmMVN}
B_t^H = \frac1{\Gamma(H+1/2)} \left\{ Z_t + \int_0^t (t-s)^{H-1/2}\,\d W_s \right\},
\end{equation}
where 
\[ Z_t = \int_{-\infty}^0 \left[ (t-s)^{H-1/2} - (-s)^{H-1/2}\right]\,\d W_s \]
has absolutely continusous trajectories and for example in order to develop a stochastic calculus with respect to this $B_t^H$, it is sufficient to consider only the Riemann-Liouville term \citep[p. 122]{Alos00}. To avoid the semi-infinite domain, \cite{Molchan69} introduced the representation \eqref{e:fBmV} with kernel
\begin{align}
K(t,s) &= C_H \Biggl[ \left( \frac{t}{s} \right)^{H-\frac12} (t-s)^{H-\frac12} 
- \left(H-\frac12\right) s^{H-\frac12} \int_s^t z^{H-\frac32} (z-s)^{H-\frac12} \d z \Biggr] \label{e:fBmMGk} \\
C_H &= \sqrt{\frac{2 H \Gamma\left(\frac32-H\right)}{\Gamma\left(H+\frac12\right)\Gamma\left(2-2H\right)} }. \notag
\end{align}
To understand the connection between the Molchan-Golosov and Mandelbrot and Van Ness representations, we refer readers to the paper by \cite{Jost08}. In particular, the Molchan-Golosov representation can be written in terms of the Gauss's hypergeometric function and it can be shown that \eqref{e:fBmMVN} is a consequence of the Molchan-Golosov representation with kernel \eqref{e:fBmMGk}. It is also worth to mention that the representation \eqref{e:fBmRL} leads to the autocovariance \eqref{e:fBmCov} with a different normalizing constant in front. To get exactly \eqref{e:fBmCov}, one needs to consider the following \emph{simplified} Riemann-Liouville representation
\begin{equation}\label{e:fBm}
B_t^H  = \sqrt{2H} \int_0^t (t-s)^{H-1/2}\,\d W_s.
\end{equation}
To avoid confusion, \eqref{e:fBm} will be the representation used in comparison of all considered numerical simulations below.

Finally, in this paper, we consider the {$\alpha$RFSV} model, firstly introduced by \cite{MerinoPospisilSobotkaSottinenVives21ijtaf}, in which the volatility process follows
\begin{equation}\label{e:aRFSV}
\sigma_t = \sigma_0\exp\left\{\xi B^H_t {- \frac12\alpha\xi^2 r(t)} \right\}, \quad t\geq 0,
\end{equation}
where $(B^H_t,t\geq 0)$ is one of the above mentioned representations of fBm and $\sigma_0>0$, $\xi>0$ and $\alpha\in[0,1]$ are model parameters together with empirical $H<1/2$. For $\alpha = 0$ we get the non-stationary RFSV model \citep{Gatheral18}, for $\alpha = 1$ the {rBergomi} model \citep{Bayer16}. Values of $\alpha$ between 0 and 1 give us a new degree of freedom that can be viewed as a weight between these two models.

Similarly to the RFSV and rBergomi models, the $\alpha$RFSV model is able to replicate the stylized facts of volatility even by using relatively small number of parameters $(\sigma_0, \xi, \rho, H, \alpha)$. However, because of the non-markovianity of the model, we cannot derive even a semi-closed-form solution using the standard It\^o calculus nor the Heston's framework. Therefore, to price a mere vanilla option, we have to rely on Monte-Carlo (MC) simulations. In addition, \cite{MerinoPospisilSobotkaSottinenVives21ijtaf} proposed a calibration scheme that combines the approximation formula derived therein alongside MC simulations.

\section{Methodology}\label{sec:methods}

In this section, we introduce the Monte-Carlo simulations methods to simulate the rough Volterra models, in particuar the $\alpha$RFSV model. Furthermore, we investigate the suitability of the so-called turbocharging variance reduction technique \citep{McCrickerd18}.

\subsection{Monte-Carlo simulations}\label{s:meth:Simul}

The price $C_t$ at time $t$ of a European call option with the strike $K$ and maturity $T$ can be expressed as
\begin{equation}\label{e:call_price}
C_t = \e^{-r(T-t)}\E_\Q \left[ (S_T-K)^+ \right], 
\end{equation}
where $\Q$ is the risk-neutral probability measure, $S_T$ is the value of the process \eqref{e:model} at time $t = T$, and $(S_T - K)^+ = \max(S_T - K, 0)$. Thus, having $M$ sample paths of the stock price process $S$ under the risk-neutral measure, 
the price of a call option \eqref{e:call_price} can be estimated by
\begin{equation}\label{e:MC_base_estim}
\hat{C}(t) = \e^{-r(T-t)}\frac{1}{M} \sum_{i=1}^M ((S_T)_i - K)^+,
\end{equation}
where $(S_T)_i$ denotes the $i$-th realization. The more sample paths we employ, the more accurate the result is. 
In fact, the rate of convergence of MC is $O( M^{-1/2})$. 

There are various methods to simulate Volterra processes but we focus on simulation of the fractional Brownian motion. We often divide these methods into two classes: exact methods and approximate methods \citep{Dieker02}. Exact methods usually exploit the covariance function of the fBm to simulate exactly the fBm (the output of the method is a sampled realization of the fBm). The advantage is obviously the exactness, however the simulation using exact methods get much slower, the more steps we simulate. For example, the Hosking method or the Cholesky method use a covariance matrix to generate the fBm from two independent normal samples. The matrix grows with every step and the calculation becomes very time and memory demanding for large samples. The second class consists of approximate methods that often use some of the integral representations of the fBm (Stochastic representation method) or they are based on the Fourier transform and its implementation fast Fourier transform (FFT), such as the spectral method \cite{Yin96}. For an extensive list of simulation methods of the fBm, see \cite{Dieker02}.

Recently, an approximate method called the Hybrid scheme introduced by \cite{Bennedsen17} has been recognized. The main idea is to discretize the stochastic integral representation of the process in the time domain and approximate the kernel function by a power function near zero and by a step function elsewhere. Later, an extension of the Hybrid scheme consisting of several variance reduction techniques was introduced by \cite{McCrickerd18}. Yet another approximation method has been proposed recently by \cite{Horvath17}. It is based on the idea of extending the Donsker's approximation of the Bm to the fBm. 

When we simulate the fBm using either exact or approximate methods, we should investigate whether the numerical samples satisfy the theoretical properties of the simulated process. In Section \ref{s:result:simul}, we compare moments estimates to the corresponding exact values. For a deeper analysis of the quality of approximate samples, see \cite{Dieker02}, Chapters 3 and 4. 

We should also mention that there are several sources of potential error \cite[Sec.~5]{Higham01}. From the the following list, the first two errors will be of the main importance for us, although the other two should not be avoided neither.
\begin{itemize}
\item \emph{Sampling error}: the error of estimation of an expected value by a sample mean. The \emph{standard MC error} of order $O(1/\sqrt{P})$, where $P$ is the number of sample paths used, can be furter reduced by considering for example quasi-random sequences.
\item \emph{Discretization error}: the error resulting from the fact that a continuous-time process is represented by a finite number $n$ of discrete-time evaluations.
\item \emph{Random number generator (RNG) error}: the bias arising from the method of generating pseudo-random numbers, the lack of independence in the samples, etc.
\item \emph{Rounding error}: the error arising from the limitations of the finite precision arithmetic, see also \cite{DanekPospisil20ijcm}.
\end{itemize}

\subsection{Simulation methods} \label{s:meth:Simul:SimMethods}

We examine three methods for simulation of the fractional Brownian motion that can be further used for simulating the $\alpha$RFSV model. We mention this because not all methods suitable for simulation of fBm paths can be used for simulation of paths of the $\alpha$RFSV model. The potential problem lies in the fact that the Bm driving the stock price process is correlated with the Bm that is used to simulate the fBm that drives the volatility process. If a method does not use the Bm to simulate the fBm, i.e., it does not operate with an integral representation of the fBm, it cannot be properly correlated with the stock price process. The example is Spectral method \cite{Yin96} that generates the fBm from a sample of uniformly distributed variables using the Fast Fourier Transform.    

The three methods, we focus on, are the Cholesky method, which is an exact (no approximation is involved) method, the Hybrid scheme, and the rDonsker scheme which are both approximate methods. We briefly describe the idea behind the algorithms in the following text.

Since we can simulate only discrete-time processes, we adapt a discrete-time notation $Y_{t_0}, Y_{t_1}, \ldots$ for the values of the fBm at the time moments $t_0, t_1, \ldots$. Once the path of the fBm is simulated for equidistant time steps, realization on another eqidistantly spaced interval is obtained by using the self-similarity property. 

\subsubsection{Cholesky method}\label{ssec:Chol}

The Cholesky method is an exact method that exploits the Cholesky decomposition of the covariance matrix of the simulated process. It means that the covariance matrix can be expressed as $L L^{\mathrm{T}}$ , where $L$ is a lower triangular matrix and $L^{\mathrm{T}}$ is its transposition. A matrix $L$ is said to be a lower triangular matrix if its elements $l_{ij} = 0$ for every $i < j$.  It can be shown that such a decomposition exists for every symmetric positive definite matrix.

If we consider a discrete realization of the fBm $Y_0, Y_1, \ldots, Y_n$, we can denote the corresponding covariance matrix $\Gamma_n$, which is a $(n+1)\times(n+1)$ matrix that can be expressed as
\begin{equation*}
\Gamma_n = 
\begin{bmatrix}
\gamma(0) &  \gamma(1) &   \gamma(2) &   \dots & \gamma(n)   \\
\gamma(1) &  \gamma(0) &   \gamma(3) &   \dots & \gamma(n-1) \\
\gamma(2) &  \gamma(3) &   \gamma(0) &   \dots & \gamma(n-2) \\
\vdots    &  \vdots    &   \vdots    &  \ddots & \vdots      \\
\gamma(n) &\gamma(n-1) & \gamma(n-2) &   \dots & \gamma(0) 
\end{bmatrix},
\end{equation*}
where $\gamma(k) = \Cov[B^H_n, B^H_{n+k}] =  \E [B^H_n B^H_{n+k}] = \frac{1}{2}[(k+1)^{2H} + (k-1)^{2H} - 2k^{2H}]$ is the autocovariance function of the discretized fBm that is derived from \eqref{e:fBmCov}. 

Since $\Gamma_n$ is a symmetric positive definite matrix, we can find its Cholesky decomposition $\Gamma_n = L_n L_n^{\mathrm{T}}$, where $L_n = (l_{i,j})_{i,j = 0}^n$. By generating a sample $v_0, \ldots, v_{n}$ from i.i.d. standard normal variables $(V_i)_{i=0}^n$, we can compute
\begin{equation*}
B_i^H = \sum_{k=0}^{n} l_{ik}v_k
\end{equation*} 
for every $i=1, \ldots, n$. Then $(0, B^H_1, \ldots, B^H_n)$ is a path of the fBm. 

For simulating $P$ paths of the fBm, the matrix notation is useful. First, generate $P$ samples from $(V_i)_{i=0}^n$ and organize the realizations into a matrix $X_n = (v_{ij})_{i=0,j=1}^{n,P}$, where $v_{ij}$ is the $j$th realization of $V_i$. Then compute
\begin{equation*}
B = L_n X_n
\end{equation*} 
and replace the first row of $B$ by zeros. Finally, notice that $B^{\mathrm{T}}$ is a $P \times (n+1)$ matrix that consists of $P$ paths of the fBm $B^H$ organized in the rows of the matrix $B^{\mathrm{T}}$.  

To practically simulate $P$ paths of $n$ steps, we firstly compute the covariance matrix $\Gamma_n$ for the required Hurst parameter $H$ in the form given above. Then, we find the Cholesky decomposition of $\Gamma_n$ by computing the lower triangular matrix $L_n$. Next generate matrix $V$ of numbers from the standard normal distribution. By computing $(L_n V_n)^{\mathrm{T}}$ we obtain $P \times (n+1)$ matrix of fBm paths organized by rows. Note that the $P \times n$ matrix of increments of the \emph{driving} Wiener process can be easily obtained as $\Delta W = \sqrt{\Delta t} V^{\mathrm{T}}$, where the partitioning $\Delta t = T/n$ is equidistant.

The price for exact simulation is, however, the time complexity of $O(n^3)$, see \cite{Asmussen07}, Chapter XI, Sect. 2.

\subsubsection{Hybrid scheme}\label{ssec:HS}

The Hybrid scheme introduced by \cite{Bennedsen17} is an approximate method that can be used for simulation of a broader class of stochastic processes called truncated Brownian semi-stationary (TBSS) processes. If $X_t$ is such a process, it can be represented as 
\begin{equation} \label{e:meth:HS:TBSS}
X_t = \int_0^t g(t-s)\nu_s \d W_s,
\end{equation}
where $W$ is the Brownian motion, $g:(0, \infty) \rightarrow [0,\infty)$ is a Borel-measurable function (a deterministic kernel function), and $\nu = \{\nu_t, t \geq 0 \}$ is a stochastic process with locally bounded trajectories that drive the volatility (intermittency) of the process. The authors list more specific assumptions on $g$ and $\nu$ in their paper to ensure that the integral \eqref{e:meth:HS:TBSS} is well defined for the whole class of TBSS processes.

In order to use the Hybrid scheme to simulate the fBm, we recall the Volterra process \eqref{e:Volterra} and 
let
\begin{equation*} \label{e:meth:HS:TBSS_fbm}
Y_t = \sqrt{2H} \int_0^t (t-s)^{H-\frac{1}{2}} \d W_s,
\end{equation*}
where $H \in (0,1)$ is the roughness parameter. The process $Y_t$ is, in fact, a TBSS process that can be obtained from \eqref{e:meth:HS:TBSS} by substituting 
\begin{equation}\label{e:gx}
g(x) = \sqrt{2H} x^{H-\frac{1}{2}},
\end{equation}
and $\nu \equiv 1$, which is exactly the Rieman-Liouville representation \eqref{e:fBm}.

The main idea behind the Hybrid scheme, as described in the abstract of the paper by \cite{Bennedsen17}, is to \emph{approximate the kernel function $g$ by a power function near zero and by a step function elsewhere. Then, the resulting approximation of the process is a combination of Wiener integrals of the power fucntion and a Riemann sum, which is why we call the method the Hybrid scheme.}

We summarize the technical part briefly and specifically only for the Volterra process \eqref{e:meth:HS:TBSS_fbm}. Let $\mathcal{G}_t^n := \{t, t-\frac{1}{n}, t-\frac{2}{n}, \ldots \}$ be the grid for the discretization of the Volterra process and let $\kappa$ be an integer greater or equal to 1. Then, the discretization of the process \eqref{e:meth:HS:TBSS_fbm} can be represented by
\begin{equation*} \label{e:fbm2_discrete}
Y_t = \sum_{k=1}^\infty \sqrt{2H} \int_{t-\frac{k}{n}}^{t-\frac{k}{n}+\frac{1}{n}} (t-s)^{H-\frac{1}{2}} \d W_s.
\end{equation*}
For \emph{``large''} $k > \kappa$, we approximate 
\begin{equation}\label{e:g_aprrox}
(t-s)^{H-\frac{1}{2}} \approx \left( \frac{b_k}{n} \right)^{H-\frac{1}{2}} , \quad t-s \in \left[ \frac{k-1}{n},\frac{k}{n} \right],
\end{equation} 
where $b_k \in [k-1,k]$. For \emph{``small''} $k \leq \kappa$, we retain the term $ (t-s)^{H-\frac{1}{2}}$ as is. Proceeding in such a way, we obtain 
\begin{equation}\label{e:HS}
Y_t \approx \sqrt{2H} \left(
\sum_{k=1}^\kappa \int_{t-\frac{k}{n}}^{t-\frac{k}{n}+\frac{1}{n}} (t-s)^{H-\frac{1}{2}}  \d W_s +
\sum_{k=\kappa+1}^\infty \left( \frac{b_k}{n} \right)^{H-\frac{1}{2}}  \int_{t-\frac{k}{n}}^{t-\frac{k}{n}+\frac{1}{n}}  \d W_s \right).
\end{equation}

To make a numerical simulation feasible, we truncate the second sum in \eqref{e:HS}. Furthermore, the values of $b_k$ that minimizes the mean square error induced by the discretization are in the form 
\begin{equation*}
b_k^* = \left( \frac{k^{H+\frac{1}{2}} - (k-1)^{H+\frac{1}{2}}}{{H+\frac{1}{2}}} \right)^{\frac1{H-\frac12}}, k\geq \kappa+1.
\end{equation*}
For derivation of the optimal $b_k^*$, see \citep[Prop.~2.8]{Bennedsen17}. Considering the truncation and the optimal $b_k^*$ for \eqref{e:HS}, we obtain the so-called Hybrid scheme.

We implement the Hybrid scheme similarly as in the \citep[Sec.~3.1]{Bennedsen17}. Suppose we simulate the process $Y$ on an equidistant grid 
$t_i = iT/n$, $i=0, 1, 2, \dots, n$, 
where $n$ is the number of steps in the interval $[0,T]$, for some $T>0$, i.e., we generate the 
discrete samples $Y_i:=Y_{t_i}$, $i=0, 1, 2, \ldots, n$.
If we consider only the first order of approximation ($\kappa = 1$), the numerical scheme for the Volterra process can be written in the form
\begin{equation}\label{e:HS_num}
Y_i = \sqrt{2H} \left(  W_{\max\{i-1,0\},1} + \sum_{k=2}^{i} \left( \frac{b_k^*}{n} \right)^{H-\frac{1}{2}} W_{i-k, 2} \right),
\end{equation}
where $W_{i,1}$ and $W_{i,2}, i = 0, 1, \ldots, n-1$ denote two random i.i.d. vectors from a bivariate normal distribution with zero mean and covariance matrix $\Sigma$ given by
\begin{equation*}
\Sigma = 
\begin{bmatrix}
\frac{1}{n} &  \frac{1}{\left( H + \frac{1}{2} \right)n^{\left( H + \frac{1}{2} \right)} } \\
 \frac{1}{\left( H + \frac{1}{2} \right)n^{\left( H + \frac{1}{2} \right)} } & \frac{1}{2Hn^{2H}} \\
\end{bmatrix}.
\end{equation*}

For the sake of the efficiency, we denote the second term on the right-hand side of \eqref{e:HS_num} by
\begin{equation*}
 \sum_{k=2}^{i} \left( \frac{b_k^*}{n} \right) W_{i-k, 2}^n =  \sum_{k=1}^{i} \Gamma_k \Xi_{i-k} = \left( \Gamma \star \Xi \right)_i, 
\end{equation*}
where $\Gamma \star \Xi $ stands for discrete convolution and
\begin{align*}
\Gamma_k &:= 
\begin{cases}
0, \quad k = 1, \\
\left( \frac{b_k^*}{n} \right)^{H-\frac{1}{2}}, \quad k = 2, \ldots, i,
\end{cases}\\
\Xi_k &:= W_{k,2}, \quad k = 0,1,\ldots, n-1.
\end{align*}
The final form of the numerical scheme is then
\begin{equation*}
Y_i = \sqrt{2H} \left(  W_{\max\{i-1,0\},1} + \left( \Gamma \star \Xi \right)_i \right) .
\end{equation*}

With respect to the recommendations of \cite{Bennedsen17} given at the bottom of page 947, our implementation of the HS uses $\kappa=1$ since the value $\kappa=2$ would be of the same quality.  
We implemented the Hybrid scheme in MATLAB and we used the Fast Fourier Transform to compute the discrete convolution. This way the complexity of the method is $O(n \log n)$, see \citep[Remark 3.2]{Bennedsen17}.

\subsubsection{rDonsker scheme}\label{ssec:rDS}

We now describe the rDonsker scheme as it was introduced by \cite{Horvath17}. Following the Section 3.3 therein, 
we update their Algorithm 3.3 for the purposes of our studied model. 

For a fixed $n\in\N$, we consider the equidistant time partition 
$t_i = iT/n$, $i=0, 1, 2, \dots, n$, of $[0,T]$ with $T>0$.
Let $Y_i^j$ denote the $j$-th discrete numerical approximation (path), $j=1,\dots,P$, of the Volterra process $Y$ evaluated at the time point $t_i$, $i=0, 1, 2, \dots, n$.

\begin{algorithm}[Simulation of the process \eqref{e:fBm} using the rDonsker scheme]\label{algo:rDonsker}~\\
\begin{enumerate}
\item Simulate two $\Nn(0,1)$ matrices
$\{\xi_{j,i}\}_{\substack{j=1,\ldots,P\\i=1,\ldots,n}}$ and
$\{\zeta_{j,i}\}_{\substack{j=1,\ldots,P\\i=1,\ldots,n}}$
with $\text{corr}(\xi_{j,i},\zeta_{j,i})=\rho$. 
Denote 
\[ \Delta W^{j}_i = \sqrt{T/n} \zeta_{i,j}, \quad i=1,\ldots,n,\quad\text{ and }\quad j=1,\ldots,P. \]
\item Simulate $P$ paths of the Volterra process $Y$ by
\[ Y^{j}_i = \sum_{k=1}^{i} g(t_{i-k+1}) \Delta W^{j}_k = \sum_{k=1}^{i} g(t_k) \Delta W^{j}_{i-k+1}, 
\quad i=1,\ldots,n,\quad\text{ and }\quad j=1,\ldots,P,
\]
where $g(x)$ is given by \eqref{e:gx}.
This step is easily implemented using discrete convolution with complexity $\Oo(n\log n)$ \citep[App. B]{Horvath17}.
\end{enumerate}
\end{algorithm}

\subsubsection{Simulation of the $\alpha$RFSV model}

Once we have simulated numerically the Volterra process $Y$ using one of the above described method, we still need to simulate the volatility process \eqref{e:volprocess} and the log-price process \eqref{e:log-model} and consequently the price process \eqref{e:model}.

Let $n\in\N$ be fixed and let us consider the equidistant time partition 
$t_i = i \Delta t$, $i=0, 1, 2, \dots, n$, of $[0,T]$ with $T>0$ and $\Delta t = T/n$. 
Let $Y_i^j$, $\sigma_i^j$, $X_i^j$ and $S_i^j$ denote the $j$-th discrete numerical approximation (path), $j=1,\dots,P$, of the Volterra process $Y$, the volatility process $\sigma$, log-stock price process $X$ and stock price process $S$ respectively, evaluated at the time point $t_i$, $i=0, 1, 2, \dots, n$.

\begin{algorithm}[Simulation of the $\alpha$RFSV model]\label{algo:aRFSV}~\\
\begin{enumerate}
\item Simulate $P$ sample paths of the Volterra process $Y_i^j$, $i=0, 1,\ldots,n$, $j=1,\ldots,P$, by one of the methods described in Sections \ref{ssec:Chol}, \ref{ssec:HS}, or \ref{ssec:rDS}. Each method produces also the increments of the driving Wiener process $\Delta W_i^j$, $i=1,\ldots,n$, $j=1,\ldots,P$.  
\item Simulate $P$ sample paths of the volatility process $\sigma$ by
\[ \sigma^{j}_i = f(t_i,Y^{j}_i), \quad i=0, 1,\ldots,n,\quad\text{ and }\quad j=1,\ldots,P, \]
where $f$ is given in \eqref{e:volprocess}.
\item Simulate $P$ sample paths of the independent Wiener process $\widetilde{W}_i^j$, $i=0, 1,\ldots,n$, $j=1,\ldots,P$, whose increments we denote by $\Delta \widetilde{W}_i^j$, $i=1,\ldots,n$, $j=1,\ldots,P$. To properly handle the correlation between the volatility process $\sigma$ and the log-price process $X$ (and consequently the price process $S$), we simulate the process $Z_i^j = \rho W_i^j + \sqrt{1-\rho^2} \widetilde{W}_i^j$, $i=0, 1,\ldots,n$, $j=1,\ldots,P$, whose increments we denote by $\Delta Z_i^j$, $i=1,\ldots,n$, $j=1,\ldots,P$.  
\item Use the forward Euler scheme to simulate $P$ sample paths of the log-price process $X$ by
\begin{align*}
\Delta X_i^j &= \left(r-\frac{1}{2} \sigma_{i-1}^j\right)\Delta t + \sqrt{\sigma_{i-1}^j}\Delta Z_i^j, 
    \quad i=1,\ldots,n,\quad\text{ and }\quad j=1,\ldots,P, \\
X_0^j &= X_0, \quad j=1,\ldots,P, \\
X_i^j &= X_0 + \sum_{k=1}^i \Delta X_k^j, 
    \quad i=1,\ldots,n,\quad\text{ and }\quad j=1,\ldots,P. \\
\end{align*}
\item Finally, we obtain the $P$ sample paths of the asset price process $S$ as 
\[ S_i^j = \exp\{X_i^j\}, \quad i=0,1,\ldots,n,\quad\text{ and }\quad j=1,\ldots,P.\]
\end{enumerate}
\end{algorithm}

It is worth to mention that in order to eliminate the RNG error mentioned at the end of subsection \ref{s:meth:Simul}, it is recommended \citep{Higham01} to have a finer discretization granularity of the driving Wiener process (or processes) than the simulated process (here $X$ or $S$ respectively). On the other hand, this brings additional memory requirements and since the RNG error is usually very small, this principle has not been applied in our numerical experiments. From the implementation point of view, steps 2-5 can be easily vectorized. The only bottleneck is step 1 that can be easily vectorized only for the Cholesky method.

\subsection{Variance reduction techniques – turbocharging} \label{s:meth:Simul:turbo}

Monte-Carlo simulations can be time demanding when we want to achieve higher precision. To further improve its efficiency, one or a combination of more variance reduction techniques can be implemented. The idea is to reduce the variance of the final estimation, and thus be able to achieve the same level of precision with smaller samples.

There are several approaches that can be used such as antithetic variates, control variates, or importance sampling. However, there is not a universal way to implement them. Instead, according to \cite{Glasserman03}, Chapter 4, p. 185, \emph{the greatest gains in efficiency from variance reduction techniques result from exploiting specific features of a problem, rather than form generic application of generic methods.} To reduce the variance of the price estimator for the $\alpha$RFSV model, we use the approach developed by \cite{McCrickerd18}, called \emph{turbocharging}. 

Pricing a call option with strike $K$ and maturity $T$ under the $\alpha$RFSV model, the idea of turbocharging is to use a \emph{mixed estimator} for the estimation of the call option price $C(t) = \e^{-r(T-t)}\E[ ( S_T - K)^+ ]$, instead of the standard MC estimator \eqref{e:MC_base_estim}. The mixed estimator is defined as
\begin{align} \label{e:mixed_est}
\begin{split}
\tilde{C(t)} &= \frac{1}{M}\sum_{i=1}^M (X_i + \hat{\omega} Y_i) - \hat{\omega} \E[Y], \\
X &= \BS \left( S_t^1,K,T,   (1-\rho^2) \int_0^t \sigma_u \d u,     r, t  \right), \\
Y &= \BS \left( S_t^1,K,T,   \rho^2 \left( \hat{Q} - \int_0^t \sigma_u \d u  \right),     r, t  \right),
\end{split}
\end{align}
where $\BS(\cdot)$ is the standard Black-Scholes formula for a call option and instead of using the $\alpha$RFSV stock price process $S_t$ \eqref{e:model} with volatility process \eqref{e:expVolterra}, it operates with its orthogonal separation into $S_t^1$ and $S_t^2$, where the process $S_t^1$ solves an SDE of the form
\begin{align}
\frac{\d S_t^1}{S_t^1} = \left( r - \frac{1}{2} \rho^2 \sigma_t \right) \d t + \rho\sqrt{\sigma}_t \d W_t.
\end{align}
Parameters $\hat{\omega}$ and $\hat{Q}$ are computed after simulation of $X$ and $Y$ as
\begin{align}
\begin{split}
\hat{\omega} &= -\frac{\sum_{i=1}^P(X_i - \hat{X})(Y_i - \hat{Y})}{\sum_{i=1}^P(Y_i - \hat{Y})^2},\\
\hat{Q}      &= \max \left\{ \left( \int_0^t \sigma_u \d u  \right)_i   : i=1,\ldots,P \right\}.
\end{split}
\end{align}

The mixed estimator \eqref{e:mixed_est} is always biased because of non-linearity of $\BS(\cdot)$. However, in  \cite{McCrickerd18}, it is stated that for $n=1000$ the bias is \emph{never practically meaningful}. We verify that in subsection \ref{s:result:simul:HSvsTurbo}. Moreover, we empirically compare variances of the standard Hybrid scheme and turbocharged Hybrid scheme and test its stability and reliability also in subsection \ref{s:result:simul:HS_var_anal}.

Ultimately, the turbocharging method is not exclusive for the Hybrid scheme. It can be successfully implemented also for the Cholesky method and for the rDonsekr scheme and used for high-precision pricing \citep{MatasPospisil21roughrobust}. 


\section{Numerical results} \label{s:result:simul}

In this section, we compare the exact Cholesky method (CM) and the approximate methods of Hybrid scheme (HS) and rDonsker scheme (rDS). We examine the quality of samples obtained from the Hybrid scheme, examine how much variance is reduced by the turbocharging technique, and we analyze the price estimation by the Hybrid scheme.

We usually choose the number of steps $n$ discretizing a path we simulate to be multiples of $252$, which is the average number of trading days in a year. For example, choosing $n = 4\times252$ for the stock price process means that the numerous daily price movements are approximated by $4$ steps of the process.

\subsection{Quality of fBm samples} 

As we mentioned earlier in Subsection \ref{s:meth:Simul}, when an approximate method is used to simulate paths of a random process, we should verify whether the generated samples possess corresponding theoretical properties. In the case of the fBm, we check whether the absolute\footnote{We consider absolute moments instead of standard moments because its visualization is more illustrative.} sample moments fit the theoretical values.

According to \citet[Remark 1.2.2.]{Mishura08}, the $q$th absolute moment of the fBm for $q \in \mathbb{N}$ can be expressed as 
\begin{equation} \label{e:fBm_qth_moment}
\E\left[ \abs{W_t}^q \right] = \frac{2^{\frac{q}{2}}}{\sqrt{\pi}} \Gamma\left( \frac{q+1}{2} \right) \abs{t}^{qH},
\end{equation}
where $\Gamma(\cdot)$ denotes the gamma function. Then, we estimate the $q$-th absolute moment \eqref{e:fBm_qth_moment} from a sample of $P$ paths by 
\begin{equation*}
\widehat{\E\left[ \abs{B^H_t}^q \right]} = \frac{1}{P} \sum_{t=1}^P (B^H_t)^q.
\end{equation*}

\begin{figure}[]
\hspace*{-2cm}\includegraphics[width=18cm]{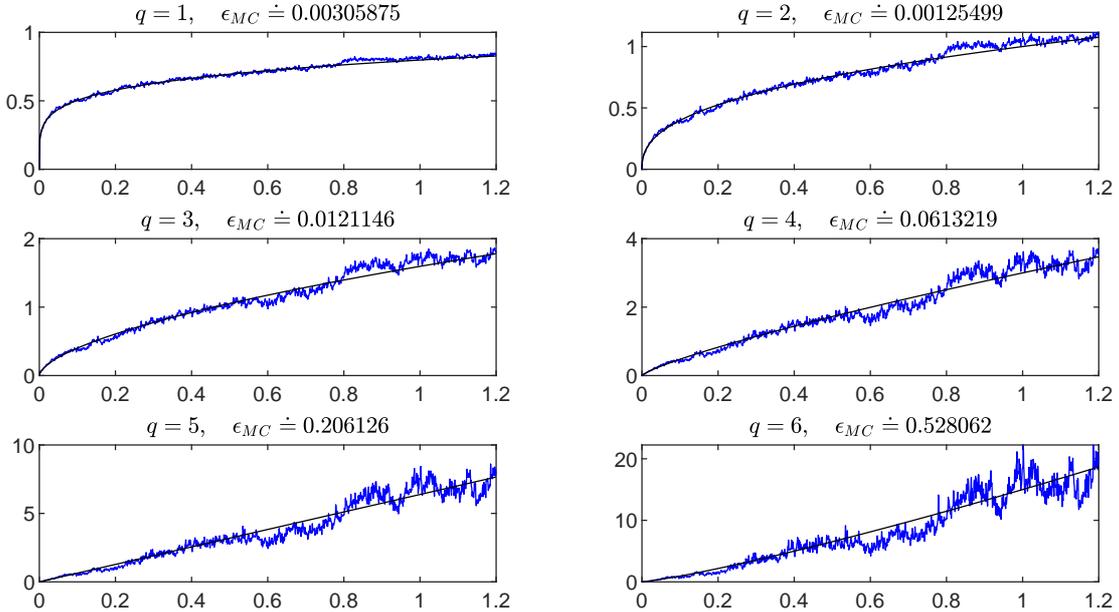}
\caption{The sample absolute moments (blue) matching the corresponding theoretical values $\E\left[\abs{B^H_t}^q \right]$ (black) given by \eqref{e:fBm_qth_moment}, of the fBm $\{ B^H_t, 0\leq t \leq 1.2 \}$ for $H = 0.20$. We simulated $P=1,000$ paths, with granularity $n=4\times 252$ steps, using the Hybrid scheme. We denote $\epsilon_{MC}$ the absolute error of the end value.}
\label{f:fBm_moments_Psmall}
\end{figure}

In Figure \ref{f:fBm_moments_Psmall}, we see an illustrative example of the sample moments fitting the corresponding theoretical values for $q = 1, \ldots, 6$. We see that with increasing value of $q$, the sample moment values diverge more from the theoretical values. Naturally, as the number of paths in the sample increases, the fit of the sample moments is closer to the theoretical values (see Figure \ref{f:fBm_moments_Pbig} in Appendix \ref{sec:A1}). 

To compare samples generated by CM, HS, and rDS methods, we calculated the distance of the absolute moments and the sample moments at the end value, similarly as in Figure \ref{f:fBm_moments_Psmall}. Since CM is an exact method, we considered it as a benchmark to the approximate HS and rDS. Additionally, we focused only on the end values since we are interested in pricing European options whose payoff depends only on the price of the stock at its maturity.

We measured the average absolute error and its variance of sample absolute moments of the fBm at the end value from 30 simulated samples, each consisting of $P=10,000$ fBm paths on $[0,1]$ with granularity $n=k\times 252, k=4, 10, 40$ steps. The results were very similar for different values of $H$. We include results for $H=0.05, 0.15, 0.40$, as $H=0.15$ is close to the empirically estimated value by \cite{Gatheral18} and we chose the other two values closer to the border values of $H$. In Table \ref{t:HSvsChol}, we summarize the results for $n = 4 \times 252$. The results for $n=k \times 252, k=10, 40$ are available in Table \ref{tab:comparison_n_10} and Table \ref{tab:comparison_n_40} (see Appendix \ref{sec:A2}) respectively. The conclusion is that based on the mentioned empirical results, both HS and rDS produce samples of very similar quality to those obtained by an exact method.

\begin{table}
\centering
\begin{tabular}{@{}lllllll@{}}
\toprule
$H=0.05$ & \multicolumn{2}{l}{Cholesky Method} & \multicolumn{2}{l}{Hybrid Scheme} & \multicolumn{2}{l}{rDonsker Scheme} \\ \midrule
q & Mean     & Variance & Mean     & Variance & Mean     & Variance \\ \hline
0 & 0.007130 & 0.000034 & 0.006829 & 0.000027 & 0.008868 & 0.000044 \\
1 & 0.004046 & 0.000010 & 0.004764 & 0.000011 & 0.004607 & 0.000014 \\
2 & 0.009212 & 0.000047 & 0.011343 & 0.000074 & 0.011072 & 0.000076 \\
3 & 0.023762 & 0.000325 & 0.028900 & 0.000513 & 0.029102 & 0.000491 \\
4 & 0.071963 & 0.002893 & 0.079237 & 0.004400 & 0.083461 & 0.003951 \\
5 & 0.239440 & 0.032847 & 0.241760 & 0.043463 & 0.257300 & 0.038733 \\
6 & 0.869100 & 0.425800 & 0.819550 & 0.501590 & 0.843800 & 0.456090 \\ \bottomrule
\end{tabular}
\begin{tabular}{@{}lllllll@{}}
\toprule
$H=0.15$ & \multicolumn{2}{l}{Cholesky Method} & \multicolumn{2}{l}{Hybrid Scheme} & \multicolumn{2}{l}{rDonsker Scheme} \\ \midrule
q & Mean     & Variance & Mean     & Variance & Mean     & Variance \\ \hline
0 & 0.007812 & 0.000034 & 0.007010 & 0.000029 & 0.009132 & 0.000042 \\
1 & 0.004422 & 0.000008 & 0.004554 & 0.000013 & 0.004343 & 0.000009 \\
2 & 0.009816 & 0.000044 & 0.011402 & 0.000085 & 0.010335 & 0.000054 \\
3 & 0.024537 & 0.000332 & 0.029753 & 0.000617 & 0.026698 & 0.000356 \\
4 & 0.070207 & 0.003052 & 0.082575 & 0.005003 & 0.078050 & 0.002622 \\
5 & 0.226630 & 0.030967 & 0.244590 & 0.045102 & 0.243630 & 0.025803 \\
6 & 0.787210 & 0.375570 & 0.791270 & 0.441700 & 0.810940 & 0.303830 \\ \bottomrule
\end{tabular}
\begin{tabular}{@{}lllllll@{}}
\toprule
$H=0.40$ & \multicolumn{2}{l}{Cholesky Method} & \multicolumn{2}{l}{Hybrid Scheme} & \multicolumn{2}{l}{rDonsker Scheme} \\ \midrule
q & Mean     & Variance & Mean     & Variance & Mean     & Variance \\ \hline
0 & 0.008177 & 0.000036 & 0.007086 & 0.000032 & 0.008921 & 0.000034 \\
1 & 0.004826 & 0.000011 & 0.004790 & 0.000011 & 0.004505 & 0.000013 \\
2 & 0.010382 & 0.000059 & 0.011527 & 0.000083 & 0.010595 & 0.000063 \\
3 & 0.024582 & 0.000362 & 0.029566 & 0.000624 & 0.027688 & 0.000353 \\
4 & 0.067798 & 0.002580 & 0.084031 & 0.005112 & 0.079051 & 0.002795 \\
5 & 0.207690 & 0.023897 & 0.264210 & 0.045314 & 0.243210 & 0.028235 \\
6 & 0.692960 & 0.277190 & 0.893930 & 0.466940 & 0.806770 & 0.324100 \\ \bottomrule
\end{tabular}
\caption{The average absolute errors (and its variances) of end values of the $q$th absolute moment of the fBm for $H = 0.05$ and for different values of $q$ calculated from 100 batches, each consisting of a sample of 10,000 fBm paths on $[0,1]$ with granularity $n=4\times 252$, generated by the Cholesky method, Hybrid scheme, and the rDonsker scheme. We can see that the samples generated by HS and rDS are very similar to the samples generated by CM. Among different values of $H$, some changes without a clear trend can be observed. For different values of $n$, no significant changes were observed (see the results in the Appendix \ref{sec:A2}).
\label{t:HSvsChol}
}
\end{table}

\subsubsection*{Runtime experiment}

Next, we compared runtimes of the CM, HS, and rDS. The asymptotic time complexity of simulation of one path is known for all of the methods we compare. Using the big $O$ notation, the time complexity of HS and rDS is $O(n \log n)$ and the time complexity of CM is $O(n^3)$. It is apparent that the approximate methods are superior to the CM considering the asymptotic complexity. However, the asymptotic behavior does not convey which method is more efficient for generating $P$ paths when $P$ is big. 

In fact, the Cholesky method is vectorized implicitly by implementing it using the matrix notation. We only compute the Cholesky decomposition of the covariance matrix of the fractional Brownian motion and apply it to an $n \times P$ matrix of normally distributed random numbers. The result after transposition is $P$ paths of the fBm organized in $P \times n$ matrix. Increasing the number of paths thus leads to the increase in the time complexity of the matrix multiplication.

On the contrary, neither the HS nor rDS can be easily vectorized. The problem is the discrete convolution of a Bm path with the convolution kernel. Therefore, the very fast computation of a fBm path has to be repeated $P$ times in a loop. Hence, the runtime increases linearly with the increasing number of paths. 

\begin{figure}[]
\includegraphics[width=\textwidth,trim=25mm 0mm 25mm 0mm,clip]{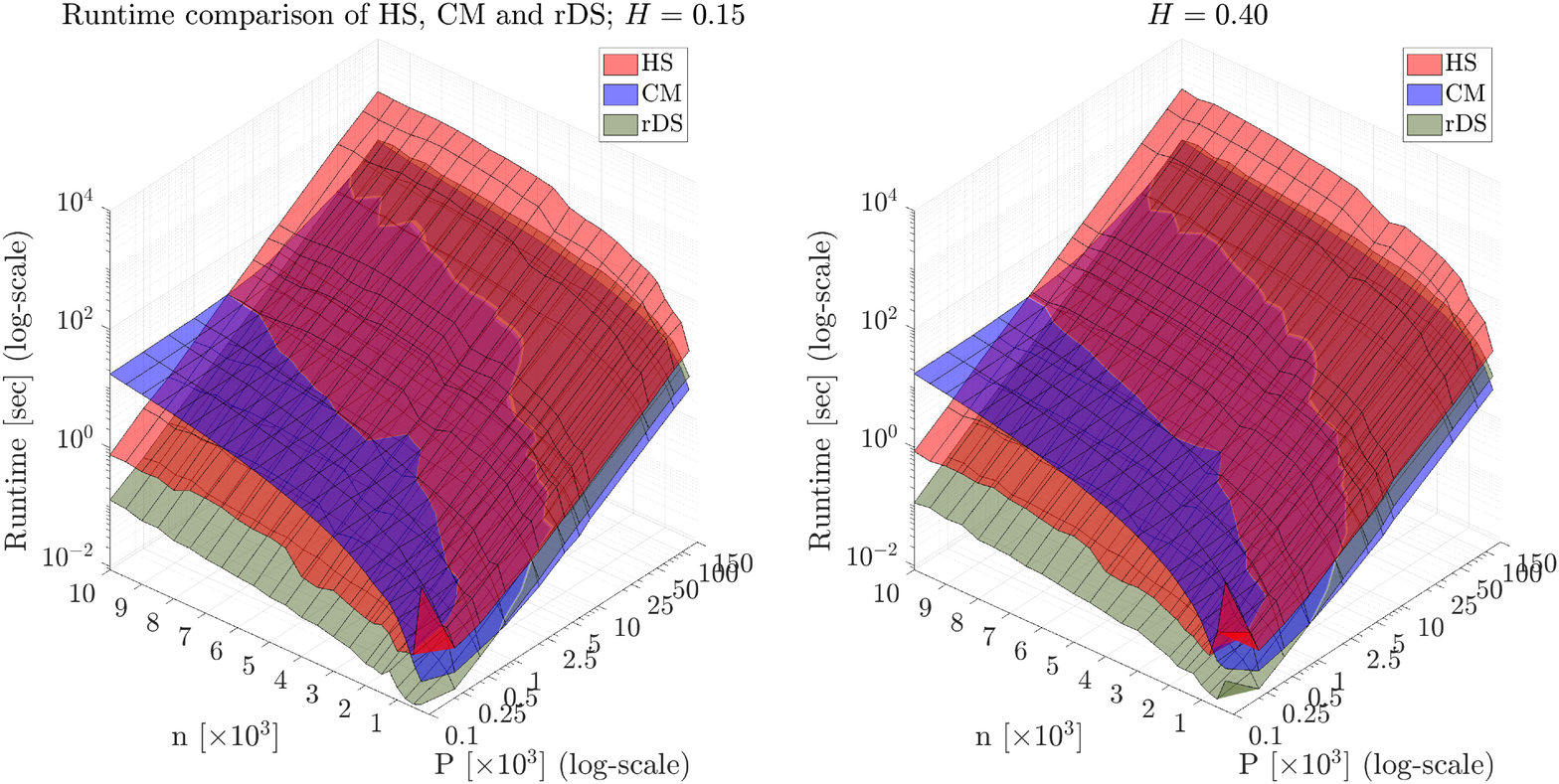}
\includegraphics[width=\textwidth,trim=25mm 0mm 25mm 0mm,clip]{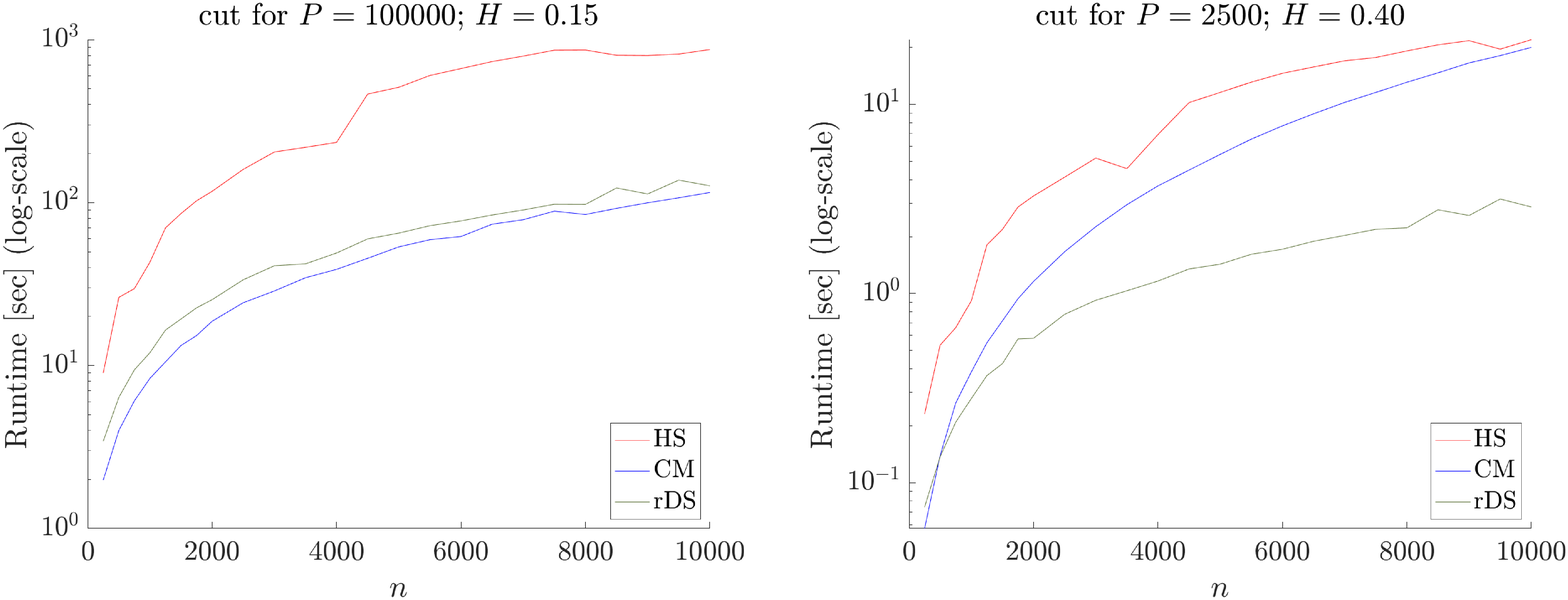}
\caption{Runtimes of the Hybrid scheme (HS), the Cholesky method (CM), and the rDonsker scheme (rDS) of simulations for different values of $P$ and $n$ on a grid and two different values of roughness parameter $H$. On top, runtimes are plotted as a surface for different values of $P$ and $n$, whereas the bottom plots show a cut of each surface for particular value of $P$.}
\label{f:runtime}
\end{figure}

To empirically compare the time efficiency of the three methods, we measured runtimes of simulations of $P$ fBm paths of $n$ steps using both methods for different values of $P$ and $n$ on a grid. Values of $P$ ranged between 100 and 150,000, values of $n$ between 250 and 10,000. The results are visualized in Figure \ref{f:runtime} where we see that while the rDS is clearly superior for small samples followed by the HS, from a certain number of paths $P$, the CM is the most efficient. The break even value of $P$ for HS is no more than 2500 for the values of $n$ we examined, that means that both CM and rDS were faster than HS for $P$ greater than 2500. For large number of paths, rDS is only slightly slower than CM. Since usually much bigger samples are necessary, the CM appears to be the best choice.

\subsubsection*{Conclusion}

Both the Hybrid scheme and the rDonsker scheme, despite being an approximate methods, generate reliable samples of fBm that are very similar to the samples obtained from an exact method - the Cholesky method. Empirically, the rDS generates a path slightly faster than HS. In fact, in the runtime experiment presented in Figure \ref{f:runtime}, HS was newer faster than rDS. The asymptotic time complexity of the rDS and the HS scheme is $O(n \log n)$ while the asymptotic time complexity of the Cholesky method is $O(n^3)$. However, when we take the number of paths into consideration, the HS and rDS are not ultimately superior to the CM in runtime, although rDS was only slightly slower than CM for large number of paths. From the experiment, whose results are visualized in Figure \ref{f:runtime}, we conclude that it is more time-efficient to simulate less paths of higher granularity by the rDS or HS while the CM is better for simulation of large number of paths.

\subsection{Simulation of the $\alpha$RFSV model} 
\label{s:res:sim_derivation_of_teor_verification}

In the previous section, we verified that the Cholesky method, the Hybrid scheme, and the rDonsker scheme produce samples of the fBm of similar preciseness and we also compared the time efficiency of the three methods. In this section, we simulate the $\alpha$RFSV model, i.e., the volatility process and the stock price process. We then analyze the effect of sample size on the pricing ability of the model and we introduce and examine the variance reduction technique turbocharging. Finally, we give recommendation on the sample size to guarantee given preciseness of the model prices.  

First, we derive formulas for the mean and the variance of the $\alpha$RFSV volatility process \eqref{e:aRFSV}. To compute the mean, consider first the exponential fBm process $\e^{\xi B^H_t}$. Since $B^H_t$ is a Gaussian process with zero mean and variance $t^{2H}$, the random variable $\e^{\xi B^H_t}$ has the log-normal distribution with mean $\e^{\frac{1}{2}\xi^2 t^{2H}}$ for every $t > 0$ and hence its $q$-th moment is
\[ \E \left[ \left( \e^{\xi B^H_t} \right)^q \right] =  \e^{\frac{1}{2} \xi^2 q^2 t^{2H}}. \]
The $q$-th moment of the volatility process \eqref{e:aRFSV} is therefore
\begin{align} 
\E \left[ \sigma_t^q \right] &= \E \left[ \sigma_0^q \exp{ \left( \xi q B^H_t - \frac{1}{2} \alpha \xi^2 q t^{2H} \right) } \right] \notag \\
                  &= \sigma_0^q \e^{ -\frac{1}{2} \alpha \xi^2 q t^{2H} } \E \left[  \e^{q \xi B^H_t}  \right]  \notag \\
                  &= \sigma_0^q\e^{ -\frac{1}{2} \alpha \xi^2 q t^{2H} }   \e^{\frac{1}{2} \xi^2 q^2 t^{2H}} \notag \\
&= \sigma_0^q \e^{ \frac{1}{2} \xi^2 q(q-\alpha) t^{2H}	} \label{e:V:q_moment}. 
\end{align}

\begin{figure}[]
\centering
\includegraphics[width=15cm]{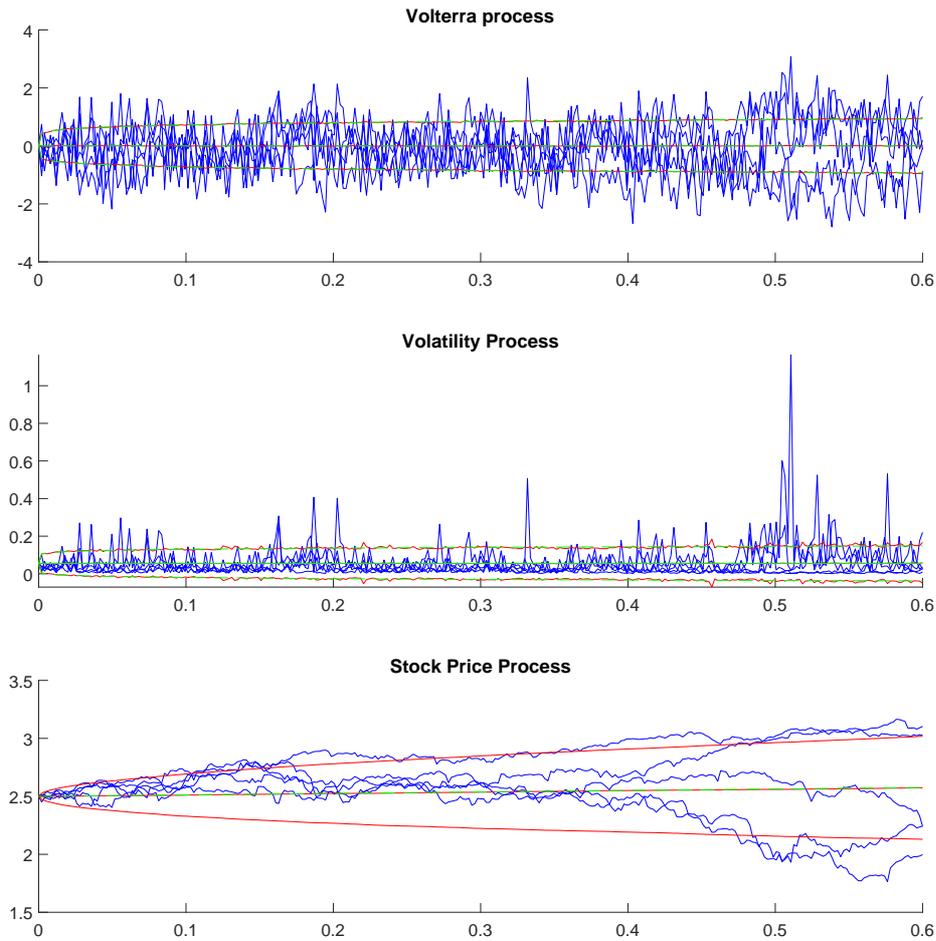}
\caption{The results of simulation of the $\alpha$RFSV model on $[0, 0.6]$ for $n=2\times252$ discretization steps and for model parameters $\alpha = 1, H = 0.07, \xi = 1.9, \rho = -0.9$, and $\sigma_0 = 0.235^2$. Five illustrative trajectories of each of the Volterra process, the volatility process, and the stock price process with $S_0 = 2.5$ and $r = 0.05$ are visualized. Also, the mean $\pm$ standard deviation for each process estimated from $P=10,000$ paths are plotted. The red curves are empirical values while the green dashed curves are the corresponding exact values. For the stock price process, only the empirical values for the standard deviation are depicted.}
\label{f:aRFSV_simulace}
\end{figure}

For illustration, we simulated $P = 10,000$ paths of the $\alpha$RFSV model on $[0, 0.6]$ for $n=2\times252$ discretization steps per unit interval using the Cholesky method for model parameters $\alpha = 1, H = 0.07, \xi = 1.9, \rho = -0.9$, and $\sigma_0 = 0.235^2$ that had been shown by \cite{Bayer16} to be consistent with the market data set used in the paper. We set $S_0 = 2.5$ and $r=0.05$ for the stock price process. The results are visualized in Figure \ref{f:aRFSV_simulace}. In each plot, five illustrative trajectories are visualized together with the theoretical mean and standard deviation (green dashed) and sample mean and standard deviation (red). We can see that for each process, the estimated values are aligned with the corresponding theoretical values. 

\subsection{Comparison of the Hybrid scheme and its turbocharged version} \label{s:result:simul:HSvsTurbo}

To quantify the effect of the turbocharging technique explained in Subsection \ref{s:meth:Simul:turbo} on the  variance of the option price estimation, we use the variance reduction factor that is simply a ratio of the variance of the prices obtained using the method with the variance reduction implemented and the variance obtained by the standard method. Its reciprocal value thus convey how many times the variance in estimated prices using the turbocharged method is smaller than the variance in estimated prices using a standard method. In this simulation study, we use the Hybrid scheme since the turbocharging was introduced in a paper where the HS was utilized, although the turbocharging can be applied to other methods as well including the Cholesky method and rDonsker scheme. Since the cases, where the turbocharging variance reduction does not behave very well is the same for all three methods, in hte rest of this section we consider only the Hybrid scheme.

We performed analyses of variance reduction by calculating the sample variances of the price obtained from the standard HS and from the turbocharged HS for different parameters. We ran simulations of $P$ paths, discretized by $n$ steps per year, in 30 batches and for each batch we calculated the prices of options for different strikes $K$. 

\begin{figure}[h!]
\centering
\includegraphics[width=13cm]{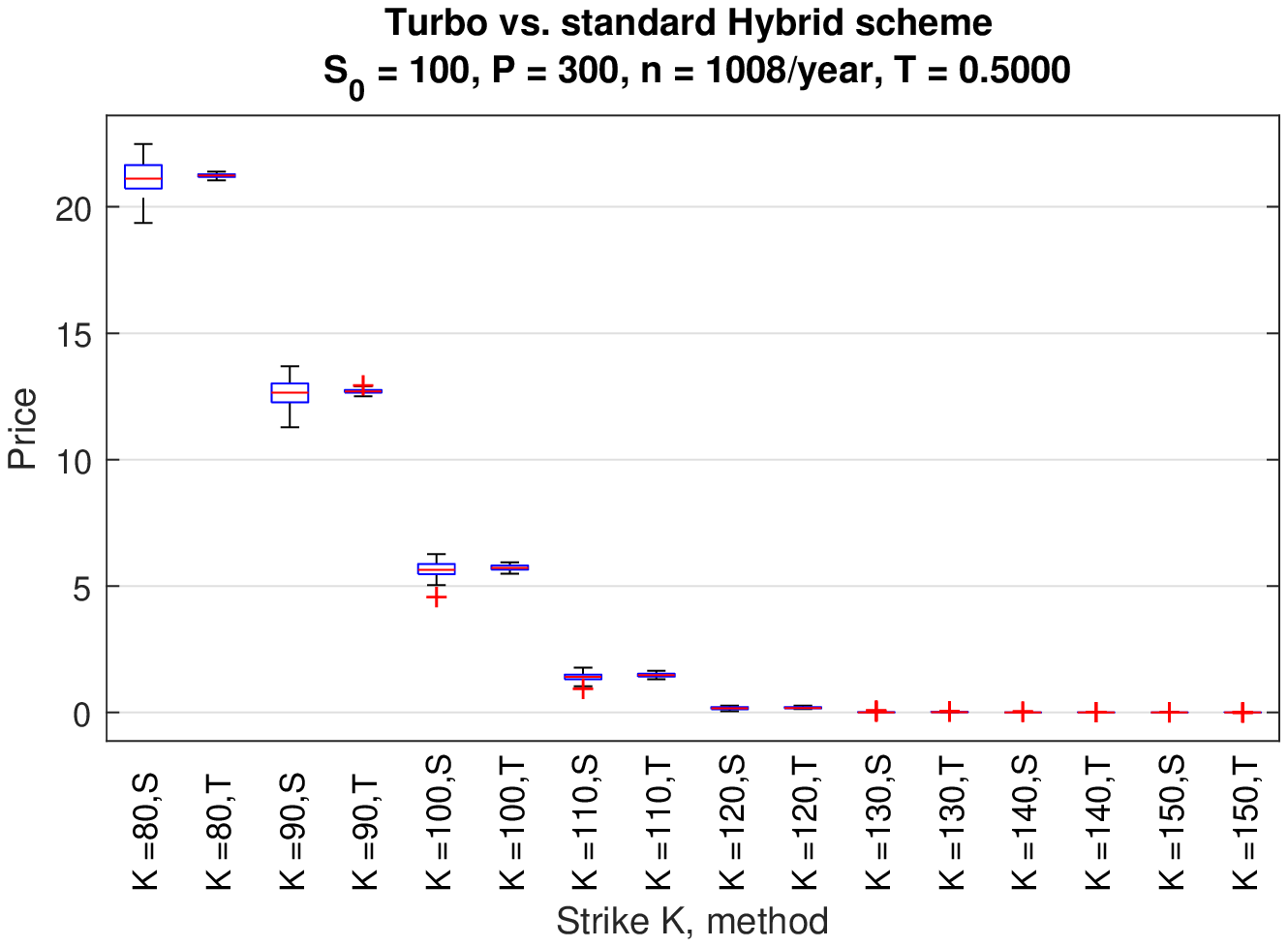}
\caption{The visualization of variance reduction from utilizing the turbocharged Hybrid scheme (T) compared to the standard Hybrid scheme (S). We simulated the $\alpha$RFSV model for $\alpha = 1, \sigma_0 = 0.235^2, H = 0.07, \rho = -0.9$, and $ \xi = 1.9$. We generated 30 batches of $P = 300$ paths, discretized by $4 \times 252$ steps, for each strike price $K$ for both methods. The boxplots depict the resulting 30 price estimates.}
\label{f:HSvsTurbo}
\end{figure}

For example, In Figure \ref{f:HSvsTurbo}, we can see how the variance is reduced while the mean does not change significantly. We chose $P=300$ paths with $n = 4 \times 252$ steps per year and model parameters that are consistent with the SPX index \cite{Bayer16}. The variance was reduced approximately $65,27,10,5,3,3,11,9$ times for the strikes $80,90, \ldots, 150$ respectively. The spot price $S_0$ was set to 100.

\begin{figure}[h!]
\centering
\includegraphics[width=13cm]{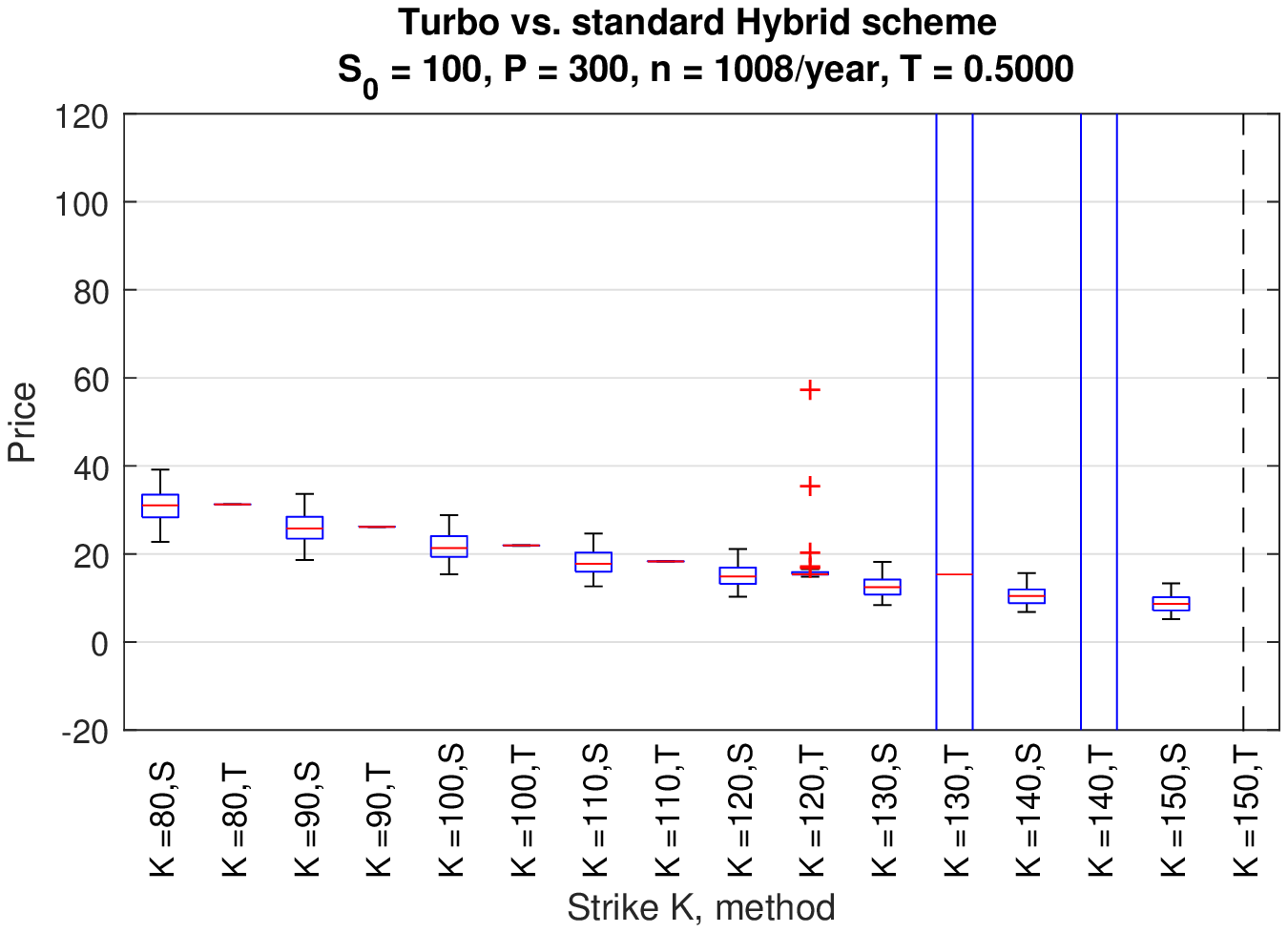}
\caption{The visualization of variance reduction from utilizing the turbocharged Hybrid scheme (T) compared to the standard Hybrid scheme (S). Coefficients of the $\alpha$RFSV model were set to $\alpha = 1, \sigma_0 = 0.62, H = 0.22, \rho = -0.05, \xi = 0.18$. We generated 30 batches of $P = 300$ paths discretized by $4 \times 252$ steps for each strike price $K$ for both methods. The boxplots depict the resulting 30 price estimates. It is apparent that in this case, the turbocharging does perform worse than the standard technique for options that are deep OTM.}
\label{f:HSvsTurbo_bad}
\end{figure}

However, the turbocharging method does not always perform well. For example, choosing $\alpha = 1, \sigma_0 = 0.62, H = 0.22, \rho = -0.05, \xi = 0.18$, and setting the other parameters as before, the resulting proce estimates are not concentrated around its mean. In fact, the variance is not reduced but extended and moreover, a bias in the estimation becomes more evident. In Figure \ref{f:HSvsTurbo_bad}, we can see that the more an option is OTM, the more the prices are scattered or completely wrong. Nonetheless, even more extreme results can be obtained for different choice of parameters. 

Based on the findings that the turbocharged HS is not always stable, we further analyzed variance reduction factors for different combinations of coefficients and parameters in order to determine, what is the source of the malfunction of the turbocharging and what is the effect of different parameters on variance reduction.

To systematically analyze and compare the quality of the option price estimates using different simulation schemes, we considered fixed spot price $S_0 = 1$ and varying model coefficients $\sigma_0, \xi, \rho, H$, the risk-free rate $r$ and the maturity $T$, all uniformly drawn from applicable intervals. For each combination of the varying coefficients, we generated 30 batches, each comprised of a sample of $P=1,000$ paths, discretized by $n = 4 \times 252$ steps per year. From each batch, we then calculated prices using the turbo HS and the standard HS for strikes $K = 0.5,0.6,\ldots,1.6$. Finally, we calculated the means and variances of prices obtained from the turbo HS and the standard HS.

We use the variance reduction factor defined as
\begin{equation} \label{e:varRedCoeff}
\frac{ \Var C^\mathcal{T} }{ \Var C^\mathcal{S} },
\end{equation}
where $C^\mathcal{T}$ is the price estimated using the turbocharged HS and $C^\mathcal{S}$ is the price estimated via the standard HS. If the variance reduction factor is negative, variance is being reduced by employing the turbocharging technique. If the factor is positive, the opposite is true.

\begin{figure}[h!]
\centering
\includegraphics[width=15cm]{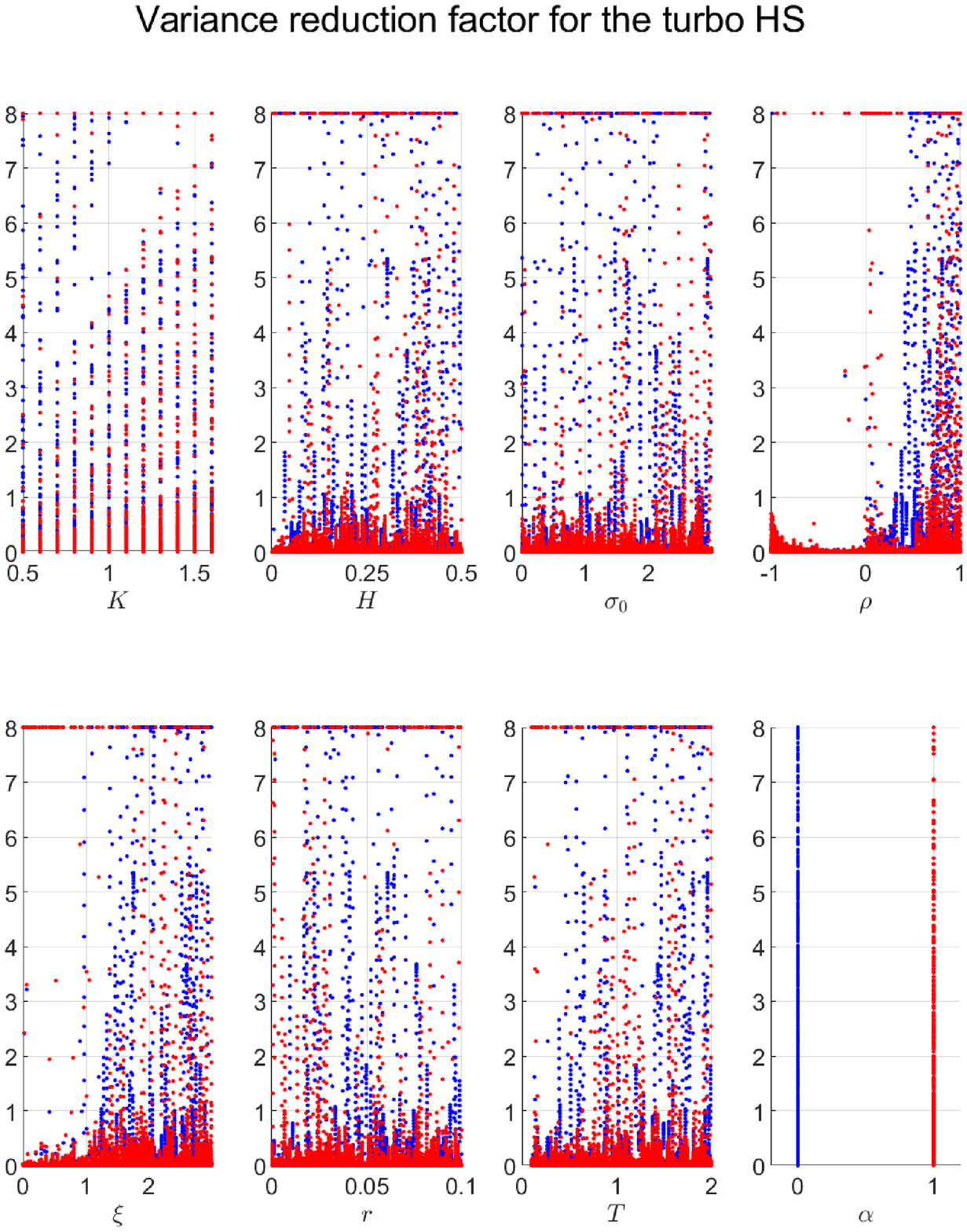}
\caption{Each point represents the variance reduction factor \eqref{e:varRedCoeff} depicted on vertical axes computed for different combinations of the model parameters that correspond to each facet (also $\alpha = 0$ corresponds to RFSV (blue) and $\alpha = 1$ corresponds to rBergomi (red)). To calculate each data point, 30 batches, each with $P=1,000$ paths discretized by $4 \times 252$ steps per $[0,1]$ were simulated.}
\label{f:varRed}
\end{figure}

The resulting variance reduction factors are depicted in Figure \ref{f:varRed} where we can observe that the amount of variance reduced clearly depends the most on $\rho$. Turbocharging appears to be the most effective for $-1 < \rho < 0 $. For $\rho \approx 0$ and $\rho \approx -1$, the variance is still being reduced except for a few outliers. Nonetheless, for $\rho > 0$, the variance is not reduced in general, i.e., the variance reduction factor is greater than 1, in more than 10\% of the cases. We can also see some dependence on $\xi$. For $\xi > 1$, the turbocharging becomes less stable. Last but not least, the coefficient $\alpha$ also appears to affect the variance reduction. For $\alpha = 1$, the turbocharging is more stable than for $\alpha = 0$. 

That leads us to conclude that for pricing a European option, we recommend using the turbocharging method only for $\rho \leq 0$. Otherwise, for $\rho > 0$, there is a significant chance, that the obtained price estimates will be very far from the true prices. In the following section we analyze the variation and bias of the estimates in a similar way. 

\subsubsection{Analysis of quality of price estimation using the turbocharged Hybrid scheme}\label{s:result:simul:HS_var_anal}

Now, we analyze the effect of turbocharging on variability and bias of model prices. To measure the variability, we use the coefficient of variation
\begin{equation}\label{e:results:simul:variance_analysis:coeff_of_variation}
\frac{\sqrt{\frac{1}{N-1} \sum_{i=1}^N \left( C_i - C \right)^2 }}{C},
\end{equation}
where $C_i$ is the estimated price from the $i$th batch and
\[ 
C = \frac{1}{P} \sum_{i=1}^N  C_i
\]
is the sample mean of the price estimates across $P$ simulated paths.

We analyzed the bias of the prices obtained using the turbo HS by comparing it to the prices obtained by the standard HS. To measure bias, we use the absolute relative error
\begin{equation} \label{e:results:simul:variance_analysis:abs_rel_err}
\frac{ \abs{C^\mathcal{T} - C^\mathcal{S} } } {C^{\mathcal{S}^{\!\!\color{white}2}}},
\end{equation}
where $C^\mathcal{S}$ is the average of price estimates obtained by the standard HS and $C^\mathcal{T}$ is the average of price estimates obtained by the turbocharged HS
 
\begin{figure}
\centering
\includegraphics[width=16cm]{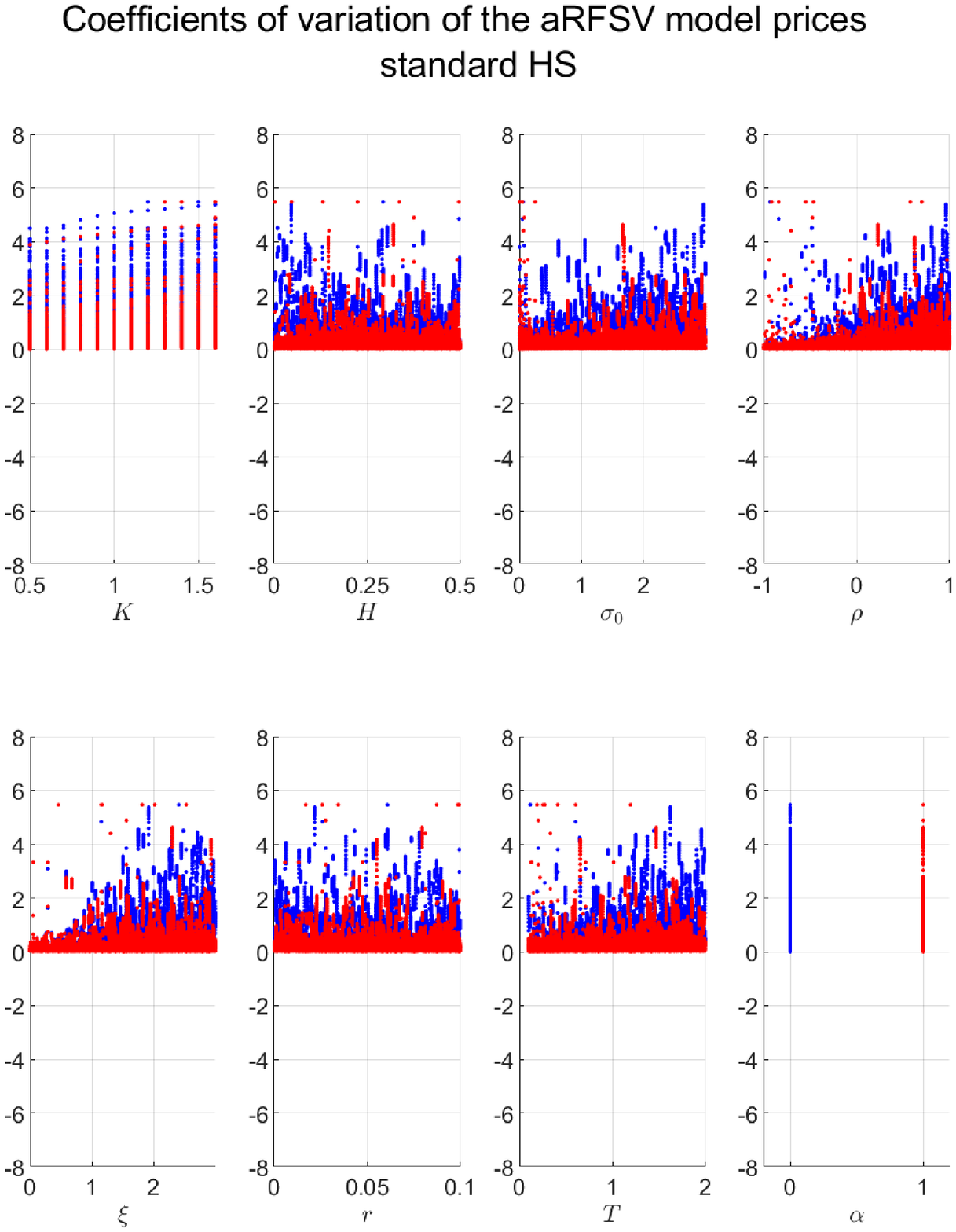}
\caption{Each point represents the coefficient of variation of price estimates calculated using the standard HS. Each coefficient of variation was computed for different combinations of the model parameters that correspond to each facet (also $\alpha = 0$ corresponds to RFSV (blue) and $\alpha = 1$ corresponds to rBergomi (red)). To calculate each data point, 30 batches, each with $P=1,000$ paths discretized by $4 \times 252$ steps per $[0,1]$ were simulated.}
\label{f:varAnalysis:CoV_off}
\end{figure}

\begin{figure}
\centering
\includegraphics[width=16cm]{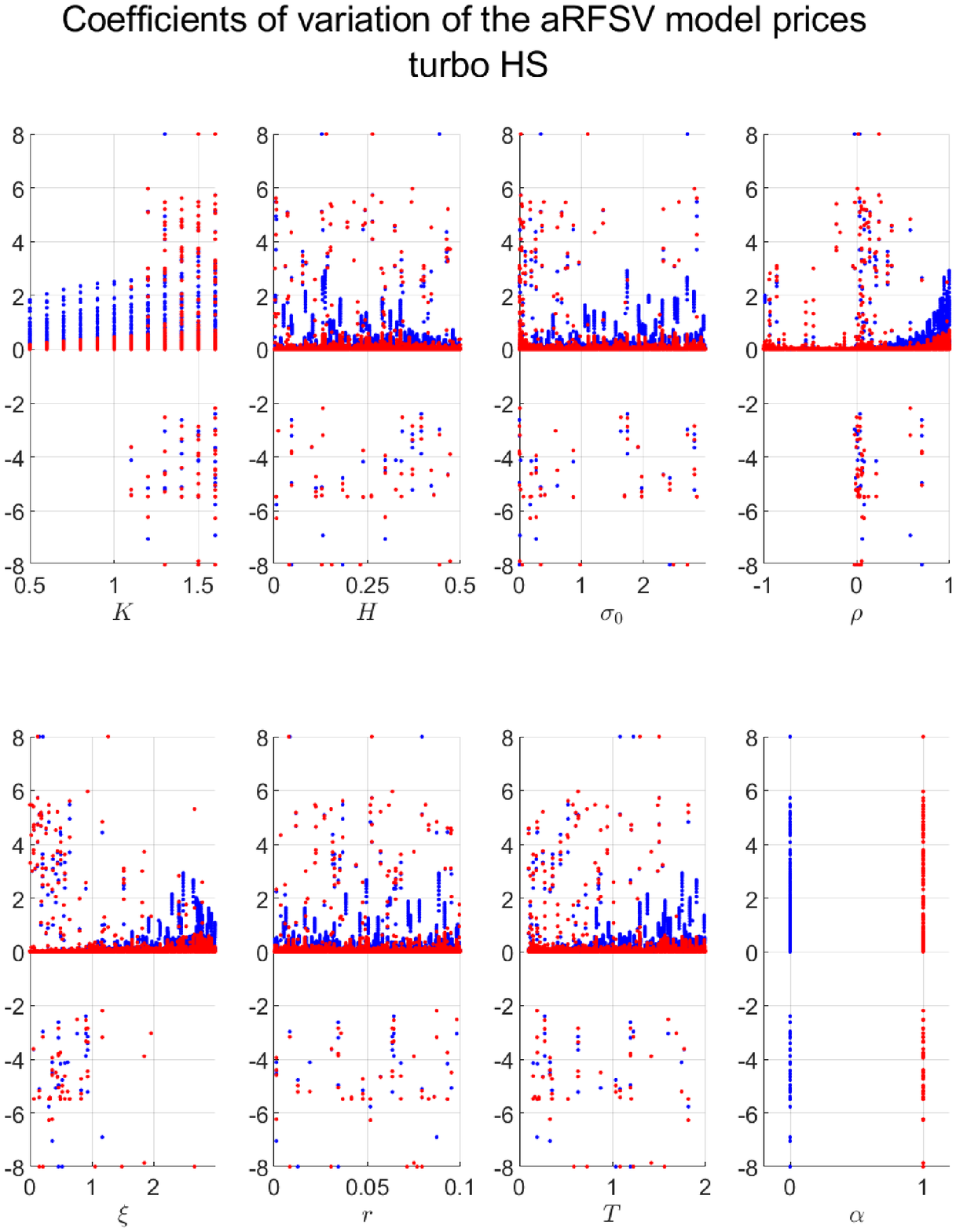}
\caption{Each point represents the coefficient of variation of price estimates calculated using the turbocharged HS. Each coefficient of variation was computed for different combinations of the model parameters that correspond to each facet (also $\alpha = 0$ corresponds to RFSV (blue) and $\alpha = 1$ corresponds to rBergomi (red)). To calculate each data point, 30 batches, each with $P=1,000$ paths discretized by $4 \times 252$ steps per $[0,1]$ were simulated.}
\label{f:varAnalysis:CoV_on}
\end{figure}

To compare the turbocharged HS estimates with the standard HS estimates, we conducted a similar simulation study as for the variance reduction factors. For different combinations of model parameters, we simulated 30 batches each comprised of $P=1,000$ paths discretized by $4 \times 252$ steps per $[0.1]$ and we calculated the coefficients of variations. We did this first using the standard HS and then the turbocharged HS. The resulting coefficients of variations are visualized in Figures \ref{f:varAnalysis:CoV_off} (standard HS) and \ref{f:varAnalysis:CoV_on} (turbocharged HS). Comparing the two figures, we see that the variance was significantly reduced for most of the parameters combinations when turbocharging was employed. However, some parameter combinations resulted in coefficients of variation being negative when turbocharging is used. A negative value of coefficient of variation can occur only when the average price $\frac{1}{N} \sum_{i=1}^N  C_i$ is negative. This leads us believe that the negative prices are a result of a malfunction of the turbocharging method. As \ref{f:varAnalysis:CoV_on} shows, the most problematic cases arose for $\rho \approx 0$ and $\xi < 1$. It also appears that only OTM options ($K>1$) are prone the turbocharging technique malfuction.

\begin{figure}
\centering
\includegraphics[width=16cm]{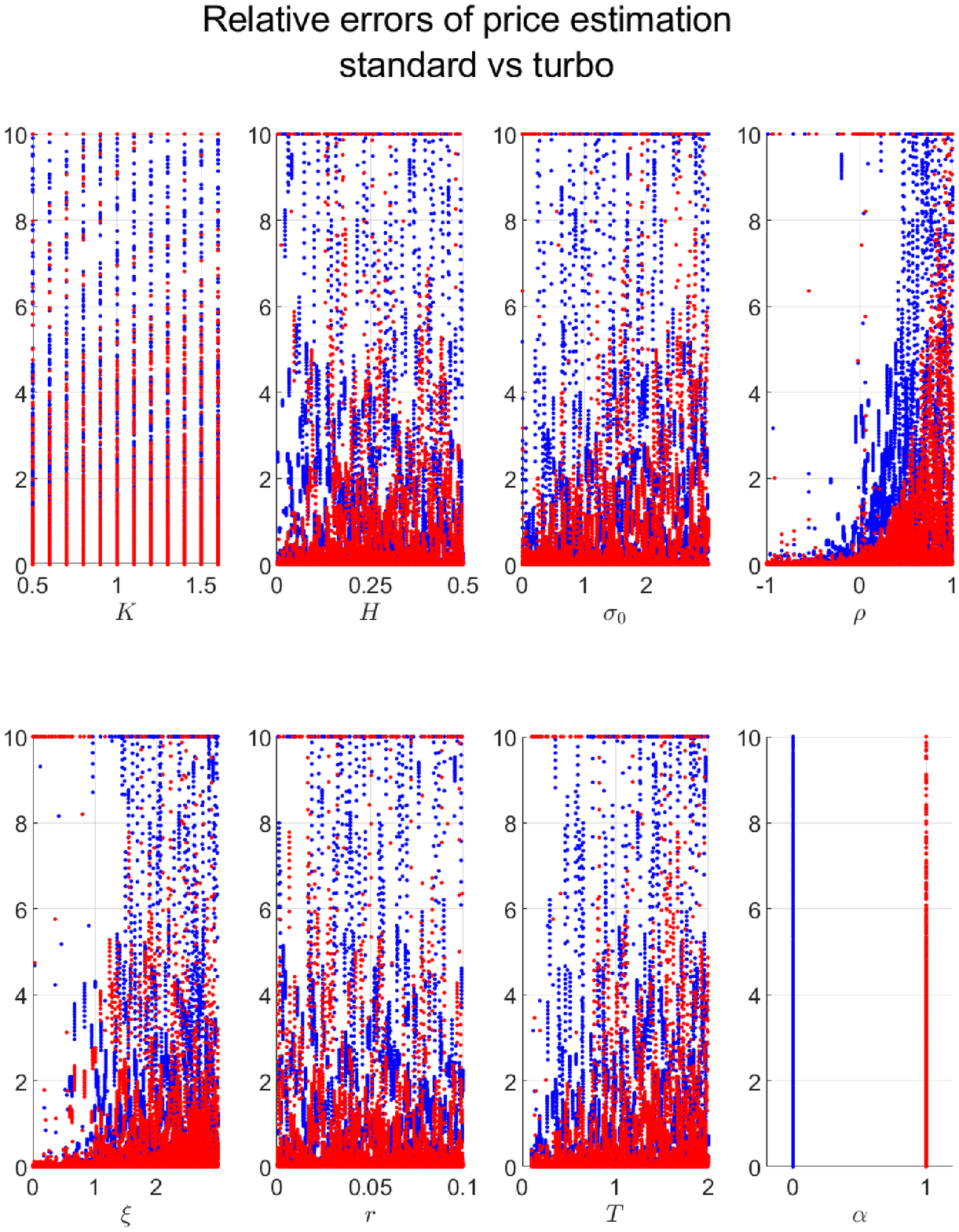}
\caption{Results of the bias analysis for the rBergomi model $\alpha = 1$ (red) and the RFSV model $\alpha = 0$ (blue). Each point depicts the absolute relative error \eqref{e:results:simul:variance_analysis:abs_rel_err} computed from 30 batches for the given combination of the model parameters. In each batch, we simulated $P=1,000$ paths, discretized by $4 \times 252$ points per $[0,1]$, using the turbo HS and the standard HS.}
\label{f:varAnalysis:CoV_price_rel_err_onVsOff}
\end{figure}

The relative errors are plotted in Figure \ref{f:varAnalysis:CoV_price_rel_err_onVsOff}. Again, we can see that  problems occur when $\rho > 0$, for which the estimation using turbocharging is apparently strongly biased compared to the estimation using the standard HS. We also see better results for $\alpha = 1$ than for $\alpha = 0$. 

\subsubsection{Modified turbocharging and the Hybrid scheme} \label{s:res:simul:modif_turbo_HS}

As a solution to the issues described earlier, we propose a modification of the turbocharging method. It is rather a naive approach that tries to identify the cases when the turbocharging does not work properly and replace the incorrect estimates by the original estimates (not turbocharged). To identify that the price of a call option is estimated incorrectly, we propose three natural criteria:
\begin{enumerate}
\item[1)] the price estimate is non-negative,
\item[2)] the price estimate is greater than the spot price of the underlying,
\item[3)] the price estimates of options with the same maturity are in descending order for increasing strike prices.
\end{enumerate} 
If at least one of those criteria is violated, suspicious price estimations are replaced by estimations using only the standard pricing method without the turbocharging. Suspicious prices are considered all prices that violates the first or the second criteria. When the third criterion is violated, having options with a given maturity $T$, the first price of an option with the strike $K_0$ that violates the descending order is considered suspicious and all prices of options with maturity $T$ and strike $K>K_0$ are considered suspicious. This way, we reduce the number and intensity of outliers among price estimates and it is guaranteed that the price estimates are non-negative and lesser than the spot price. The advantage of this approach is that it can be easily implemented and it is time-efficient – the pricing is not slowed down. 

\begin{figure}[h!]
\centering
\includegraphics[width=13cm]{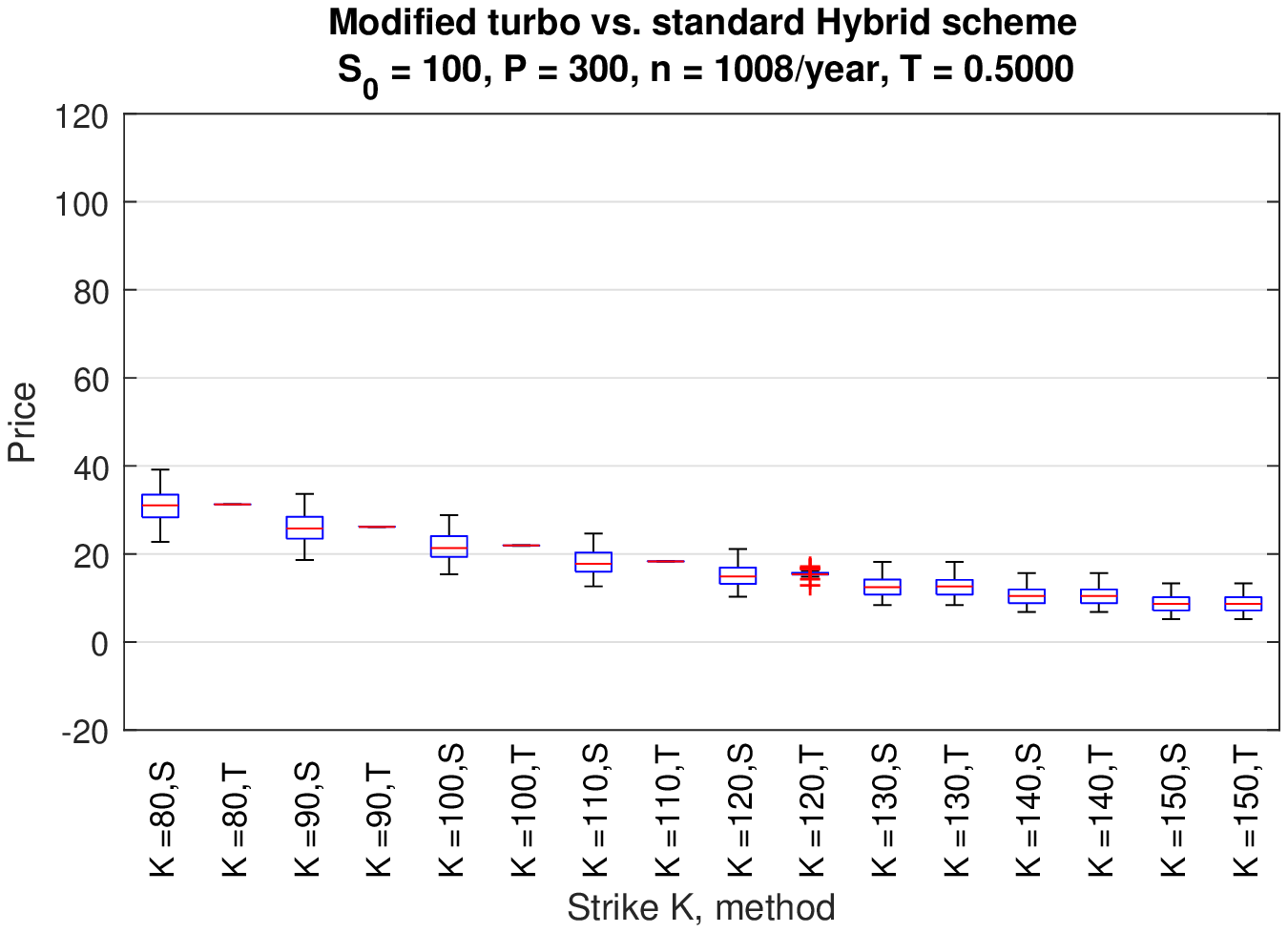}
\caption{Variance reduction visualization for the modified turbocharged Hybrid scheme w.r.t the standard Hybrid scheme (S). Coefficients of the $\alpha$RFSV model are set to $\alpha = 1, \sigma_0 = 0.62, H = 0.22, \rho = -0.05, \xi = 0.18$. We generated 30 batches of $P = 300$ paths discretized by $4 \times 252$ steps for each strike price $K$ for both methods. The boxplots depict the resulting price estimations. Compare with Figure \ref{f:HSvsTurbo_bad}.}
\label{f:HSvsModifiedTurbo}
\end{figure}

For illustration, we consider the same coefficients and parameters as for the example of the turbo HS malfunction in Figure \ref{f:HSvsTurbo_bad} but we use the modified turbocharging instead. We visualized the results in Figure \ref{f:HSvsModifiedTurbo}. In that case, modification was activated for the prices of option with $K=120$ and all following options with strike $K>120$. All those suspicious prices were replaced by the price estimates using the standard HS.

\begin{figure}
\centering
\includegraphics[width=14cm]{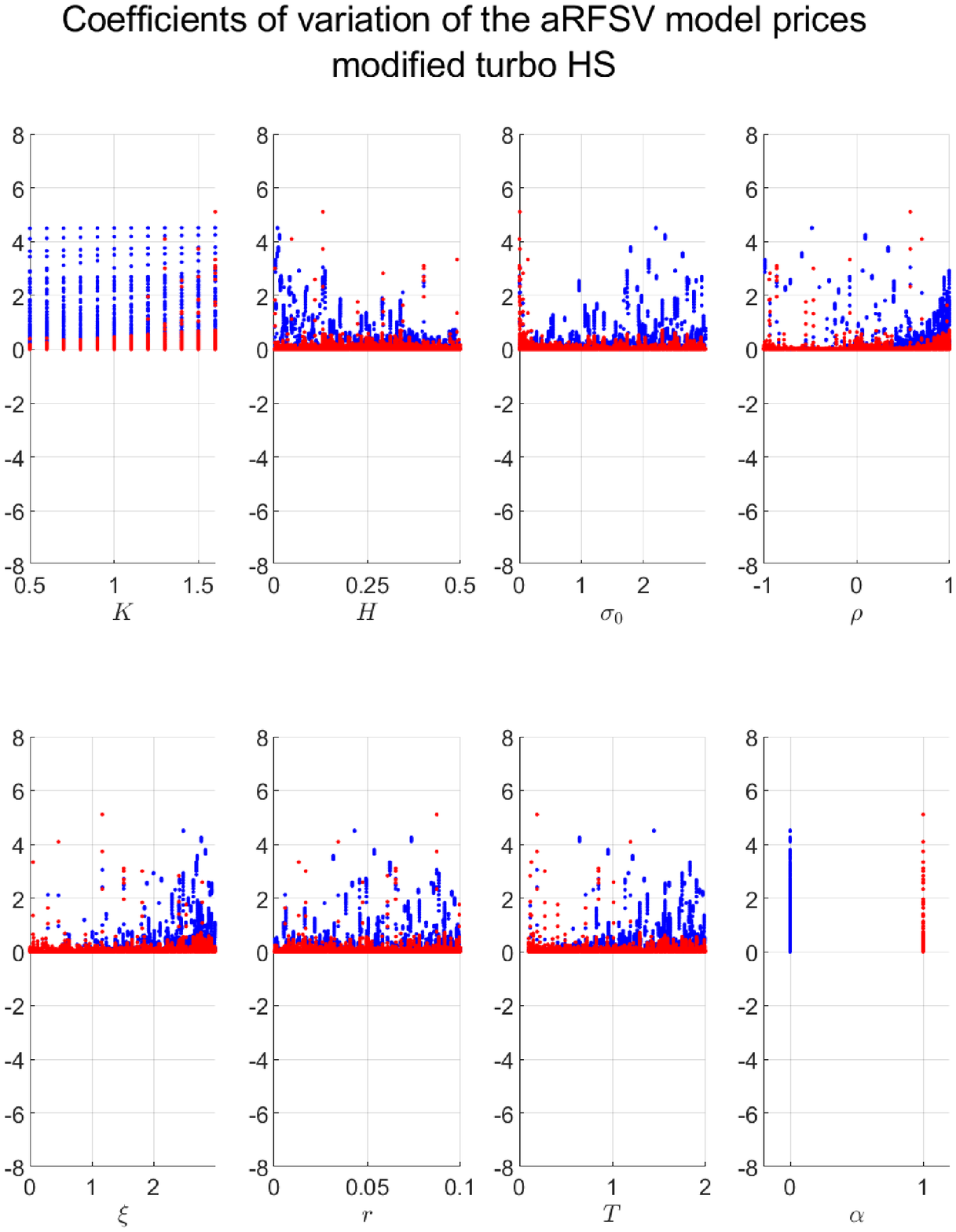}
\caption{Each point represents the coefficient of variation of price estimates calculated using the modified turbocharged HS. Each coefficient of variation was computed for different combinations of the model parameters that correspond to each facet (also $\alpha = 0$ corresponds to RFSV (blue) and $\alpha = 1$ corresponds to rBergomi (red)). To calculate each data point, 30 batches, each with $P=1,000$ paths discretized by $4 \times 252$ steps per $[0,1]$ were simulated.}
\label{f:varAnalysis:CoV_onSafe}
\end{figure}

To test the modification, we ran identical simulation as before. The results are visualized in Figure \ref{f:varAnalysis:CoV_onSafe}. We see that the modification eliminated all the negative price estimates and also majority of outlying positive estimates, especially for $\alpha = 1$. Moreover, we can observe that almost all the outlying values are now concentrated around $\sigma_0 \approx 0$. For $\alpha = 0$. In general it appears that the turbocharging works better for $\alpha = 1$. 

However, even though we managed to reduce variance and eliminate majority of cases where excessive variance can be an issue, we have to remember that for $\rho > 0$ the estimation can be strongly biased. We tested the bias of modified turbo HS but we ended up with similar results as for the turbo HS that are visualized in Figure \ref{f:varAnalysis:CoV_price_rel_err_onVsOff}. 

To sum this part up, we found out that the turbocharging method is not stable for every combination of parameters and model coefficients, especially for $\rho > 0$. We proposed a simple modification but we still recommend to avoid using turbocharging when $\rho > 0$ due to strong bias in the estimation. In fact, we suggest using the turbocharging technique only for $\rho < -0.05$. Also, we recommend avoiding turbocharging for $\sigma_0 \approx 0$ and $\xi > 2$ when $\alpha = 0$ due to higher variance compared to the standard HS.

\subsubsection{Number of sample paths needed}

When pricing options, we are usually concerned with the accuracy of obtained model prices. Standard Monte-Carlo simulations converge with $O(1/\sqrt{P})$, where, in our case, $P$ is the number of trajectories generated. Thus a question of how many trajectories we need to have a certain accuracy guaranteed arises. A trivial approach suggests to have the sampling error sufficiently small at a prescribed confidence level, which means to have the length of the corresponding confidence interval sufficiently small. When dealing with prices that are considered in standard currencies such as dollars or euros that count with the smallest unit to be one \emph{cent} (0.01), a \emph{sufficiently accurate price} is usually such that is determined with an error smaller than 0.005 with the 99\% confidence. Clearly, the length of the confidence interval for estimating the expectation of the stock price process by the sample mean (here the mean of the sample paths) depends on the corresponding quantile, standard deviation and the number of sample path $P$, typically in the order $O(1/\sqrt{P})$.

Empirically, classical non-fractional SV models require usually $P=10,000$ sample paths to attain this precision, measured typically ATM only. However, we empirically showed that all considered rough fractional SV models require at least $P = 100,000$. Moreover, the more OTM an option is, the wider the confidence interval. This phenomenon is influenced by the increasing sample standard deviation in all three considered simulation methods. 

When variance reduction techniques are employed, the number of trajectories required may be smaller but that depends on the combination of model parameters thus we recommend using at least $P = 100,000$ as a safe option even when reduction techniques are used. Also, we recommend using the Cholesky method since it is more time efficient compared to the HS or rDonsker scheme.

\FloatBarrier
\section{Conclusion}\label{sec:conclusion}

The main objectives of the paper were rough Volterra stochastic volatility models, in particular the $\alpha$RFSV model that covers both the RFSV and rBerhomi models, and their MC simulations. We showed that all considered methods (Cholesky method, Hybrid scheme, and rDonsker scheme) are appropriate  for the simulation of the $\alpha$RFSV model in terms of accuracy and that the rDonsker and HS is faster for small number of densely discretized paths, while the CM is more efficient for larger number of paths. Although the CM is of cubic complexity, it can get more efficient than the FFT implementation of rDS due to its vector implementation.

Further, we examined the variance reduction techniques for price estimation fulfilled by the turbocharging method applied specifically to the HS. We observed that for $\rho<0$, the variance is reduced significantly; however, for $\rho>0$, the turbocharging is not stable, and the variance is, in fact, higher in many cases.

We also analyzed variability of the price estimations for different combinations of the model coefficients. We showed that for certain combinations, the turbocharged HS produces outlying price estimates and that, in some instances, the estimates are heavily biased. In order to prevent excessive and incorrect prices from occurring, we proposed a simple modification of the method that identifies suspicious prices and replaces them with the estimates from the standard HS. Moreover, we recommended using stricter boundary condition  $\rho<-0.05$ in order to prevent malfunctions. Using the modified turbo HS for the $\alpha$RFSV model under the given condition, the variance is being reduced around 60 times on average. Moreover, the average coefficient of variation of price estimation is around $1.5 \%$, while the median around $0.005 \%$.

Writing the paper, several additional questions and issues arose. First of all we showed that none of the studied simulation methods is ``perfect'', however, we may conclude that the combination of CM and modified turbocharging variance reduction techniques is currently the most suitable for derivative pricing purposes that use MC simulations of rough Volterra processes like the studied $\alpha$RFSV model. A more detailed analysis would be required for the TBSS class of processes \citep{Bennedsen17,Bennedsen21}. Although the rough fractional Ornstein-Uhlenbeck or rough fractional Cox-Ingersoll-Ross processes for modelling the volatility dynamics were not included in the presented study, it is expected to get similar results. It is also worth to mention, that other control variate variance reduction techniques might improve the simulations even further, but their investigation is still an open issue.

Only recently an approximation of the price for the $\alpha$RFSV model was proposed by \cite{MerinoPospisilSobotkaSottinenVives21ijtaf}. Promising numerical properties of the approximation allows to propose a hybrid calibration scheme which combines the approximation formula alongside MC simulations that were analyzed in this paper. Only very recently a paper by \cite{FukasawaHirano21refinement} has been published where authors proposed a refinement for the HS by reducing and reusing random numbers.

\section*{Funding}
The work was partially supported by the Czech Science Foundation (GA\v{C}R) grant no. GA18-16680S ``Rough models of fractional stochastic volatility''.

\section*{Acknowledgements}
A preprint of this article became an integral part of the Master's thesis \cite{Matas21} titled \emph{Rough fractional stochastic volatility models} that was written by Jan Matas and supervised by Jan Posp\'{\i}\v{s}il.
Our sincere gratitude goes to anonymous referees for their valuable suggestions and insightful criticism.

Computational and storage resources were supplied by the project ``e-Infrastruktura CZ'' (e-INFRA LM2018140) supported by the Ministry of Education, Youth and Sports of the Czech Republic.




\appendix
\addtocontents{toc}{\protect\setcounter{tocdepth}{0}}

\section{Moments matching the theoretical moments of fBm}\label{sec:A1}
\begin{figure}[h!]
\hspace*{-2cm}\includegraphics[width=18cm]{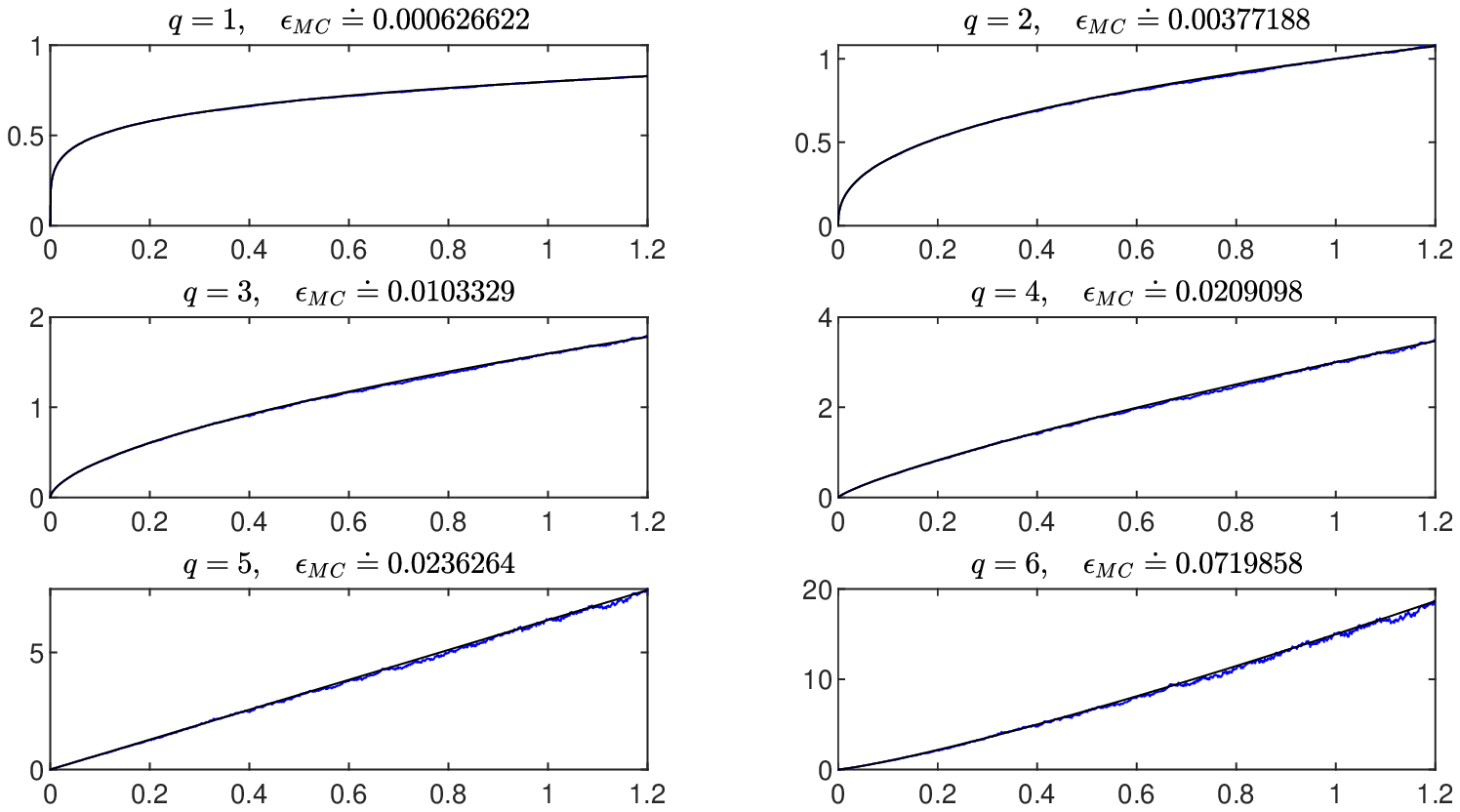}
\caption{Sample absolute moments (blue) matching the corresponding theoretical values $\E\left[\abs{B^H_t}^q \right]$ (black), see \eqref{e:fBm_qth_moment}, of fBm $\{ B^H_t, 0\leq t \leq 1.2 \}$ for $H = 0.20$. There were simulated $P=100,000$ paths, with granularity $n=4\times 252$ steps, using the Hybrid scheme. We denote $\epsilon_{MC}$ the absolute error of the end end value.
}
\label{f:fBm_moments_Pbig}
\end{figure}

\section{Moments comparison for different methods}\label{sec:A2}

\begin{table}[]
\centering
\begin{tabular}{@{}lllllll@{}}
\toprule
$H=0.05$ & \multicolumn{2}{l}{Cholesky Method} & \multicolumn{2}{l}{Hybrid Scheme} & \multicolumn{2}{l}{rDonsker Scheme} \\ \midrule
q & Mean     & Variance & Mean     & Variance & Mean     & Variance \\ \hline
0 & 0.008221 & 0.000049 & 0.008403 & 0.000034 & 0.007020 & 0.000026 \\
1 & 0.005127 & 0.000014 & 0.004572 & 0.000010 & 0.004642 & 0.000010 \\
2 & 0.011390 & 0.000073 & 0.010483 & 0.000058 & 0.010782 & 0.000063 \\
3 & 0.026252 & 0.000439 & 0.025241 & 0.000388 & 0.025893 & 0.000395 \\
4 & 0.067564 & 0.003168 & 0.070353 & 0.002973 & 0.071209 & 0.002713 \\
5 & 0.192990 & 0.028105 & 0.220440 & 0.028295 & 0.216280 & 0.025062 \\
6 & 0.624850 & 0.284910 & 0.754800 & 0.344150 & 0.722030 & 0.299070 \\
7 & 2.213200 & 3.430500 & 2.798900 & 5.018100 & 2.635200 & 4.261900 \\ \bottomrule
\end{tabular}
\begin{tabular}{@{}lllllll@{}}
\toprule
$H=0.15$ & \multicolumn{2}{l}{Cholesky Method} & \multicolumn{2}{l}{Hybrid Scheme} & \multicolumn{2}{l}{rDonsker Scheme} \\ \midrule
q & Mean     & Variance & Mean     & Variance & Mean     & Variance \\ \hline
0 & 0.008537 & 0.000035 & 0.007316 & 0.000032 & 0.007037 & 0.000024 \\
1 & 0.005085 & 0.000012 & 0.005091 & 0.000015 & 0.004659 & 0.000013 \\
2 & 0.011900 & 0.000070 & 0.011633 & 0.000087 & 0.011424 & 0.000064 \\
3 & 0.028787 & 0.000448 & 0.028371 & 0.000525 & 0.028762 & 0.000360 \\
4 & 0.076774 & 0.003740 & 0.077467 & 0.003893 & 0.079348 & 0.002799 \\
5 & 0.229880 & 0.037152 & 0.239630 & 0.033930 & 0.241770 & 0.028301 \\
6 & 0.767460 & 0.421310 & 0.814020 & 0.347690 & 0.827310 & 0.305350 \\
7 & 2.785400 & 5.447200 & 2.925800 & 4.303600 & 3.027800 & 3.874200 \\ \bottomrule
\end{tabular}
\begin{tabular}{@{}lllllll@{}}
\toprule
$H=0.40$ & \multicolumn{2}{l}{Cholesky Method} & \multicolumn{2}{l}{Hybrid Scheme} & \multicolumn{2}{l}{rDonsker Scheme} \\ \midrule
q & Mean     & Variance & Mean     & Variance & Mean     & Variance \\ \hline
0 & 0.008203 & 0.000031 & 0.007092 & 0.000027 & 0.007159 & 0.000030 \\
1 & 0.005247 & 0.000012 & 0.005146 & 0.000013 & 0.004929 & 0.000012 \\
2 & 0.011959 & 0.000061 & 0.011932 & 0.000088 & 0.011096 & 0.000067 \\
3 & 0.028778 & 0.000389 & 0.030701 & 0.000532 & 0.026263 & 0.000412 \\
4 & 0.078426 & 0.003019 & 0.086823 & 0.003866 & 0.071361 & 0.003187 \\
5 & 0.236640 & 0.029089 & 0.267050 & 0.035268 & 0.221530 & 0.030328 \\
6 & 0.793390 & 0.316990 & 0.886720 & 0.405670 & 0.752830 & 0.342010 \\
7 & 2.877600 & 4.011300 & 3.115400 & 6.029000 & 2.729000 & 4.285600 \\ \bottomrule
\end{tabular}
\caption{Moments comparison similar to Table~\ref{t:HSvsChol}, but for $n=10\times 252$.}
\label{tab:comparison_n_10}
\end{table}

\begin{table}[]
\centering
\begin{tabular}{@{}lllllll@{}}
\toprule
$H=0.05$ & \multicolumn{2}{l}{Cholesky Method} & \multicolumn{2}{l}{Hybrid Scheme} & \multicolumn{2}{l}{rDonsker Scheme} \\ \midrule
q & Mean     & Variance & Mean     & Variance & Mean     & Variance \\ \hline
0 & 0.008091 & 0.000039 & 0.008222 & 0.000031 & 0.008379 & 0.000046 \\
1 & 0.004785 & 0.000011 & 0.004503 & 0.000014 & 0.004692 & 0.000012 \\
2 & 0.011886 & 0.000064 & 0.010111 & 0.000070 & 0.010653 & 0.000081 \\
3 & 0.029170 & 0.000419 & 0.025700 & 0.000395 & 0.026156 & 0.000541 \\
4 & 0.079047 & 0.003199 & 0.071357 & 0.003143 & 0.073889 & 0.003886 \\
5 & 0.234920 & 0.030924 & 0.223220 & 0.028853 & 0.231440 & 0.033092 \\
6 & 0.763790 & 0.370130 & 0.753510 & 0.327460 & 0.780520 & 0.352530 \\
7 & 2.720800 & 5.146300 & 2.730700 & 4.372200 & 2.837000 & 4.350400 \\ \bottomrule
\end{tabular}
\begin{tabular}{@{}lllllll@{}}
\toprule
$H=0.15$ & \multicolumn{2}{l}{Cholesky Method} & \multicolumn{2}{l}{Hybrid Scheme} & \multicolumn{2}{l}{rDonsker Scheme} \\ \midrule
q & Mean     & Variance & Mean     & Variance & Mean     & Variance \\ \midrule
0 & 0.007953 & 0.000041 & 0.008069 & 0.000032 & 0.008293 & 0.000041 \\
1 & 0.004897 & 0.000013 & 0.004783 & 0.000016 & 0.004862 & 0.000012 \\
2 & 0.012491 & 0.000073 & 0.011327 & 0.000081 & 0.010363 & 0.000068 \\
3 & 0.031475 & 0.000473 & 0.028359 & 0.000417 & 0.025449 & 0.000439 \\
4 & 0.085535 & 0.003723 & 0.078515 & 0.002747 & 0.071485 & 0.003714 \\
5 & 0.257360 & 0.033203 & 0.240040 & 0.024493 & 0.229570 & 0.036517 \\
6 & 0.838530 & 0.359960 & 0.803130 & 0.277840 & 0.801930 & 0.421190 \\
7 & 2.912300 & 4.783200 & 2.859800 & 3.935500 & 2.953400 & 5.654900 \\ \bottomrule
\end{tabular}
\begin{tabular}{@{}lllllll@{}}
\toprule
$H=0.40$ & \multicolumn{2}{l}{Cholesky Method} & \multicolumn{2}{l}{Hybrid Scheme} & \multicolumn{2}{l}{rDonsker Scheme} \\ \midrule
q & Mean     & Variance & Mean     & Variance & Mean     & Variance \\ \midrule
0 & 0.007752 & 0.000045 & 0.008133 & 0.000032 & 0.007224 & 0.000039 \\
1 & 0.005078 & 0.000015 & 0.005544 & 0.000018 & 0.004427 & 0.000010 \\
2 & 0.012430 & 0.000089 & 0.013020 & 0.000084 & 0.010211 & 0.000060 \\
3 & 0.031103 & 0.000552 & 0.031506 & 0.000445 & 0.025433 & 0.000390 \\
4 & 0.084518 & 0.004132 & 0.084357 & 0.003047 & 0.070828 & 0.003046 \\
5 & 0.254270 & 0.037795 & 0.244130 & 0.029531 & 0.216880 & 0.030261 \\
6 & 0.836550 & 0.424440 & 0.775790 & 0.349170 & 0.740520 & 0.339040 \\
7 & 3.004100 & 5.390100 & 2.681400 & 4.973900 & 2.719200 & 4.354000 \\ \bottomrule
\end{tabular}
\caption{Moments comparison similar to Table~\ref{t:HSvsChol}, but for $n=40\times 252$.}
\label{tab:comparison_n_40}
\end{table}


\end{document}